\documentclass[reprint, amsfonts, amssymb, amsmath,  showkeys, pra, superscriptaddress, nofootinbib, onecolumn, longbibliography]{revtex4-2}

\usepackage{amsmath,amssymb,amsfonts,amsthm,makeidx,graphicx}
\usepackage{txfonts}
\usepackage{helvet}
\usepackage{braket}
\usepackage{bbm}
\usepackage[colorlinks=true,citecolor=blue,linkcolor=magenta]{hyperref}
\usepackage{quantikz}
\usepackage{enumitem}

\usepackage[most]{tcolorbox}

\definecolor{fundamental}{RGB}{55, 110, 111}

\newtcolorbox[auto counter]{pabox}[2][]{fonttitle=\bfseries\small,
title={#2},#1,colframe=gray}
\newtcolorbox[use counter from=pabox]{mybox}[2][]{
floatplacement=h!,
colback=fundamental!5!white,colframe=fundamental!75!black,title=#2,#1,breakable,fontupper=\small ,fontlower=\small, fonttitle=\bfseries\small}

\begin{document}

\title{A Primer on Quantum Machine Learning}

\author{Su Yeon Chang}
\affiliation{Theoretical Division, Los Alamos National Laboratory, Los Alamos, New Mexico 87545, USA}

\author{M. Cerezo}
\thanks{cerezo@lanl.gov}
\affiliation{Information Sciences, Los Alamos National Laboratory, Los Alamos, New Mexico 87545, USA}
\affiliation{Quantum Science Center, Oak Ridge, TN 37931, USA}

\begin{abstract}
Quantum machine learning (QML) is a computational paradigm that seeks to apply quantum-mechanical resources to solve learning problems. As such, the goal of this framework is to leverage quantum processors to tackle optimization, supervised, unsupervised and reinforcement learning, and generative modeling--among other tasks--more efficiently than classical models. Here we offer a high level overview of QML, focusing on settings where the quantum device is the primary learning or data generating unit. We outline the field’s tensions between practicality and guarantees, access models and speedups, and classical baselines and claimed quantum advantages--flagging where evidence is strong, where it is conditional or still lacking, and where open questions remain. By shedding light on these nuances and debates, we aim to provide a friendly map of the QML landscape so that the reader can judge when--and under what assumptions--quantum approaches may offer real benefits.
\end{abstract}
\maketitle

Given the theoretical promise of quantum computing and the empirical, often heuristic, success of classical machine learning (ML), the emergence of quantum machine learning (QML) was perhaps inevitable. 
On the one hand, quantum computers can solve tasks--such as simulating quantum systems, integer factorization, and discrete logarithms--with exponentially fewer computational resources than classical devices\footnote{Assuming, of course, that BQP (the class of decision problems solvable by a quantum computer in polynomial time) is not in P (the class of decision problems that a classical computer can solve in polynomial time)}~\cite{nielsen2000quantum}.  On the other hand, classical ML has delivered striking practical gains across vision~\cite{he2016deep}, language~\cite{brown2020language}, and scientific endeavors  more broadly, despite limited worst-case guarantees: heuristic optimizers, overparametrized models, inductive architectural biases, and aggressive scaling routinely outperform what current theory would predict. As such, QML arises from the quest to combine quantum algorithmic techniques with the pragmatic, data-driven methods of modern ML--leading to hopes of accelerating existing learning tasks and enabling new ones, especially in regimes native to quantum data.

 To contextualize this review, it is important to note that, despite its current widespread attention, QML is not a new field. The quantum perceptron was introduced in 1994~\cite{lewenstein1994quantum}, a quantum version of the probably approximately correct (PAC) learning formalism was developed in 1995~\cite{bshouty1995learning}, and initial ideas for quantum artificial neural networks (NNs) emerged in the mid-1990s~\cite{kak1995quantum,vlasov1997quantum}. Today, QML has been proposed for virtually all areas of learning, including supervised learning~\cite{lloyd2013quantum,schuld2018supervised,schuld2019quantum,havlivcek2019supervised,caro2021generalization,caro2022outofdistribution}, unsupervised ~\cite{otterbach2017unsupervised,kerenidis2019q,andreassen2019junipr,kyriienko2022unsupervised},  reinforcement ~\cite{mckiernan2019automated,saggio2021experimental,skolik2021quantum,meyer2022survey,chen2022variational}, and generative modeling~\cite{perdomo2018opportunities,benedetti2019generative,coyle2020born,sweke2021on,anschuetz2021critical,gao2022enhancing,rudolph2022generation,rudolph2023trainability}. Moreover, potential application domains include combinatorial optimization~\cite{farhi2014quantum,hadfield2019quantum,harrigan2021quantum,lee2021towards,liu2021layer,amaro2022filtering,abbas2024challenges}, quantum chemistry~\cite{peruzzo2014variational,tang2019qubit,grimsley2019adaptive,grimsley2022adapt,anastasiou2024tetris,fedorov2022vqe}, high-energy physics~\cite{guan2021quantum,wu2021application,wu2022challenges,humble2022snowmass} and finance~\cite{orus2019quantum,egger2020quantum,alcazar2020classical,herman2022survey,cherrat2023quantum}. Notably, after more than three decades, the true capabilities of QML are still unknown, and  
there is active debate about the extent to which current progress reflects intrinsic computational power versus enthusiasm and positive bias generated by combining two highly visible fields: ML and quantum computing. As it stands, there are ongoing tensions: for quantum data, QML has measurably improved how experiments extract information~\cite{huang2020predicting,huang2021quantum,king2024triply}; for classical data, promising heuristics~\cite{cerezo2020variationalreview,bharti2021noisy,peral2024systematic} coexist with no-go theorems for worst- or average-case performance~\cite{bittel2021training,larocca2024review,ciliberto2020statistical},  separations between classical and quantum learners are known for contrived tasks~\cite{servedio2001quantum,servedio2004equivalences,jerbi2023shadows,liu2021rigorous}, yet other purported advantages for practical problems that once raised considerable hope~\cite{rebentrost2014quantum,lloyd2014quantum,kerenidis2016quantum} have been dequantized~\cite{tang2019quantum,tang2021quantum,gilyen2018quantum}. As such, QML finds itself at a pivotal, actively debated stage, and the jury is still out on questions such as: \textit{For what types of data is QML well suited?  How should we empirically compare classical and quantum learning algorithms?  What (genuine) quantum mechanisms underlie potential gains over classical methods on practical scenarios?}

\begin{figure}[t]
\centering
\includegraphics[width=.9\textwidth]{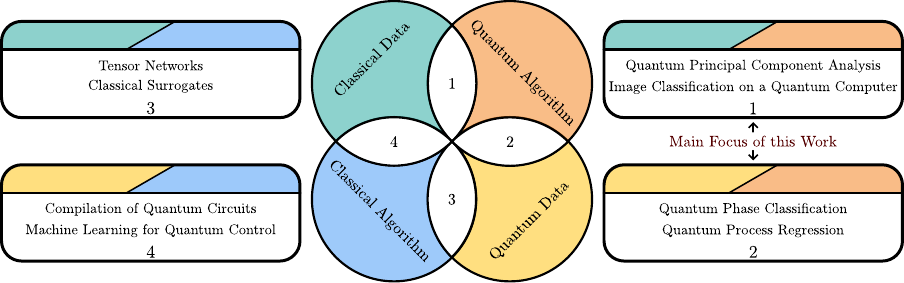}
\caption{Quantum machine learning (QML) is somewhat of an umbrella term spanning many settings. This figure schematically organizes the landscape along two axes—data type (classical vs.\ quantum) and algorithm type (classical vs.\ quantum)--yielding four illustrative categories. Examples in each box are indicative rather than exhaustive, and the boundaries are guidelines rather than strict rules as hybrid paradigms can straddle multiple categories. This review concentrates on schemes where the quantum device is the primary learning and considers both classical and quantum data (categories $1$ and $2$, respectively). }
\label{chapQML:fig1}
\end{figure}

The goal of this review is to provide a cool-headed, bird’s-eye overview of different approaches to QML, reviewing past and current methods for constructing quantum learning algorithms and highlighting both their advantages and limitations. In particular, we will illustrate the principles of QML through the lens of supervised learning, return later to unsupervised, reinforcement, an generative settings, as well as give examples of how QML has impacted other areas of research. Indeed, as illustrated in Fig.~\ref{chapQML:fig1}, QML spans many settings. Hence, throughout this article we will focus on models where the quantum device is used as the central information-processing or data-generating unit, i.e., schemes based on coherent state preparation, evolution, and measurements. Where relevant, we will distinguish expectation-value estimation from sample-based generative tasks, as the latter can lead to different resource trade-offs and potential advantages. In addition, we will be explicit about data-access assumptions throughout, as they critically shape both positive results and dequantizations. Finally, we will focus on modern and widely used approaches to QML, while only briefly mentioning techniques that have either fallen out of favor or are in their earliest stages of development. Hence, this article does not intend to be a comprehensive review, nor will it cover all topics in a balanced fashion. For additional material, we refer the reader to Refs.~\cite{schuld2015introduction,mcclean2016theory,biamonte2017quantum,dunjko2018machine}, which cover the initial stages of QML, as well as to the book Ref.~\cite{schuld2021supervised}. For more modern reviews we point to~\cite{cerezo2020variationalreview,bharti2021noisy,peral2024systematic,dunjko2020non,cerezo2022challenges}, and in particular to the recent reviews~\cite{gujju2024quantum,wang2024comprehensive,anshu2024survey,gebhart2022learning}.

\section{Learning and learning on quantum computers}\label{chapQML:sec:learning}

Before motivating the use of quantum computers for learning, it is useful to define what it means to “learn from data.” As such, this section first (briefly) introduces the PAC framework for supervised learning and then presents a simple toy problem, solved with standard classical techniques, as a stepping stone toward their quantum counterparts.

\subsection{The theory behind learning}\label{chapQML:subsec:PAC}

The key idea in ML is to use data to train a model that makes predictions or decisions without being explicitly programmed to do so~\cite{goodfellow2016deep,mohri2018foundations}. Unlike traditional programming--which takes a set of rules plus a set of inputs to produce an output (e.g., the rules of multiplication plus two integers yield their product)--the ML framework takes inputs plus outputs and produces the set of rules that connects them (e.g., pictures of cats and dogs plus their associated labels yield an algorithm for cat/dog image classification). In this context, the defining properties of a learning model are (i) how we model the connection between input and output (the hypothesis class and its inductive biases) and (ii) how we formalize learning and generalization.

While ``learning'' and its limits can be studied within several mathematical frameworks, one of the most common ones within computational learning theory is PAC learning, introduced in~\cite{valiant1984theory} (see also~\cite{mohri2018foundations,shalev2014understanding,kearns1994introduction,caro2022quantum} and the lecture notes in~\cite{de2019quantum}), and the generalizations within Vapnik–Chervonenkis (VC) theory~\cite{vapnik2013nature}. In the PAC formalism,  let $\mathcal{X}$ denote the input domain and $\mathcal{Y}$ the label set. The goal is to learn an unknown target function, or concept, $f:\mathcal{X}\rightarrow\mathcal{Y}$ from examples of the form $(x,f(x))$. In the supervised binary classification cat-dog example, $\mathcal{X}$ is the set of such images and $\mathcal{Y}=\{0,1\}$ (e.g., $0$ = cat, $1$ = dog). The standard assumption is that the training examples are drawn independently and identically distributed (i.i.d.) from a distribution $\mathcal{D}$ over $\mathcal{X}\times\mathcal{Y}$ which models the way in which the learner obtains examples by interaction with its environment. 

Next, one fixes a hypothesis class $\mathcal{HC}\subseteq\{h:\mathcal{X}\rightarrow\mathcal{Y}\}$, whose choice reflects our inductive biases (e.g., fully connected neural networks with a bounded number of layers and hidden neurons per layer). A learning algorithm seeks to find the best hypothesis $h$ within $\mathcal{HC}$. During training, we seek to reduce as much as possible its error, defined as how much the predictions of $h$ differ from those of the true, target function $f$. Typically, one can quantify how well $h$ is doing by defining a loss function $\ell: \mathcal{Y}\times\mathcal{Y}\rightarrow\mathbb{R}_{\geq 0}$.  Denoting the true label as $y=f(x)$, typical choices include the $0-1$ loss $\ell(h(x),y)=1-\delta_{h(x),y}$, the squared loss $\ell(h(x),y)=(h(x)-y)^2$ and the absolute loss $\ell(h(x),y)=|h(x)-y|$. Given a training set $\mathcal{S}=\{(x_i,y_i)\}_{i=1}^N$ drawn from $\mathcal{D}$, the empirical loss (or the training error) is\footnote{In the QML literature, the empirical loss of Eq.~\eqref{eq:emp-loss} is oftentimes simply called the loss function.}
\begin{equation}\label{eq:emp-loss}
\widehat{R}_{\mathcal{S}}(h)=\frac{1}{N}\sum_{i=1}^N \ell\big(h(x_i),y_i\big)\,.
\end{equation}

While $\widehat{R}_{\mathcal{S}}$ measures performance on the dataset, one primarily cares about how well $h$ performs on new, unseen data. For instance, one can obtain $\widehat{R}_{\mathcal{S}}(h)=0$ by a simple look-up table (which, for instance, assigns random label to data points not in $\mathcal{S}$),  indicating that  performance in ML is only truly quantified with the model's performance beyond the training examples. As such, the generalization error (also known as the true or population risk) is defined as the expected loss over $\mathcal{X}\times\mathcal{Y}$ on fresh data sampled according to $\mathcal{D}$
\begin{equation}\label{eq:pop-risk}
R _{\mathcal{D}} (h)=\mathbb{E}_{(x,y)\sim\mathcal{D}} [\ell(h(x),y)]\,.
\end{equation}
Equations~\eqref{eq:emp-loss} and~\eqref{eq:pop-risk} are related through the so-called generalization gap 
\begin{equation}\label{eq:gen-gap}
\mathrm{gap}_{\mathcal{S}}(h)=R_{\mathcal{D}}(h)-\widehat{R}_{\mathcal{S}}(h)\,.
\end{equation}
A small gap implies that the model should perform equally well on new data as it did on its training dataset. 

From the previous, training can be viewed as attempting to solve the optimization problem
\begin{equation}\label{eq:train-prob}
\min_{h\in\mathcal{HC}} \ \widehat{R}_{\mathcal{S}}(h)\,.
\end{equation}
Finding a global minimizer of Eq.~\eqref{eq:train-prob} can be difficult, and matters are further complicated when training is stochastic (e.g., different random initializations of a neural network can yield different values of $\widehat{R}_{\mathcal{S}}$ even with the same optimizer). As such, the PAC goal is to produce a hypothesis that is probably approximately correct. This leads to the definition of an $(\varepsilon,\delta)$-PAC learner. Specifically, an $(\varepsilon,\delta)$-PAC learner for a hypothesis class $\mathcal{HC}$ with respect to $\mathcal{D}$ is an algorithm that, given $N$ i.i.d.\ samples, outputs a hypothesis $h$ such that, with probability at least $1-\delta$,
\[
R_{\mathcal{D}}(h)\ \le\ \inf_{g\in\mathcal{HC}} R_{\mathcal{D}}(g)\ +\ \varepsilon\;,
\]
for some $\varepsilon > 0$.
In the realizable case (there exists $h^\star\in\mathcal{HC}$ with $R_{\mathcal{D}}(h^\star)=0$), this reduces to $R_{\mathcal{D}}(h)\le \varepsilon$.

\noindent The specific mathematical dependence of  $N$ with $\varepsilon$ and $\delta$ is known as the sample complexity. Here, one says that a problem is tractable, or that it can be efficiently solved, if there exists a learning algorithm for which the sample complexity is at most polynomial in  $1/\varepsilon$ and $1/\delta$.  

Within this context, VC theory formalizes the intuition that, while very flexible hypothesis classes (such as those containing a look-up table) can perfectly fit the training data, this says little about test performance. On the other hand, for more restricted classes, a small training error  is a reliable proxy for a small generalization error. This motivates the definition of the VC dimension $\mathrm{VC}(\mathcal{HC})=d$ as a capacity measure for the model. Explicitly, $d$ is the largest integer such that there exists a set $S\subseteq\mathcal{X}$ with $|S|=d$ that is shattered by $\mathcal{HC}$ (i.e., every labeling of $S$ is realizable by some $h\in\mathcal{HC}$). For binary classification, one of the variants of the VC theorem states that with probability at least $1-\delta$ over the draw of $\mathcal{S}$,
\[
\sup_{h\in\mathcal{HC}}\big|R_{\mathcal{D}}(h)-\widehat{R}_{\mathcal{S}}(h)\big|
\;\le\;
\varepsilon
\;:=\;
\sqrt{\frac{\,d\!\left(\log\frac{2N}{d}+1\right)}{N}+\frac{\log\frac{4}{\delta}}{N}}\,.
\]
Equivalently, for all $h\in\mathcal{HC}$,
$R_{\mathcal{D}}(h)\le \widehat{R}_{\mathcal{S}}(h)+\varepsilon(d,N,\delta)$. This important result directly ties generalization to the model complexity $d$ and the sample size $N$, and implies that to guarantee low generalization error from low training error, the model must have low complexity\footnote{In a way, this constitutes a No Free Lunch result! Models which are too flexible and perform well on a large variety of tasks are expected to have  low performance on each given problem.}.

\subsection{A toy classification problem and classical baselines}\label{chapQML:subsec:toy}

\begin{figure}[t]
\centering
\includegraphics[width=1\textwidth]{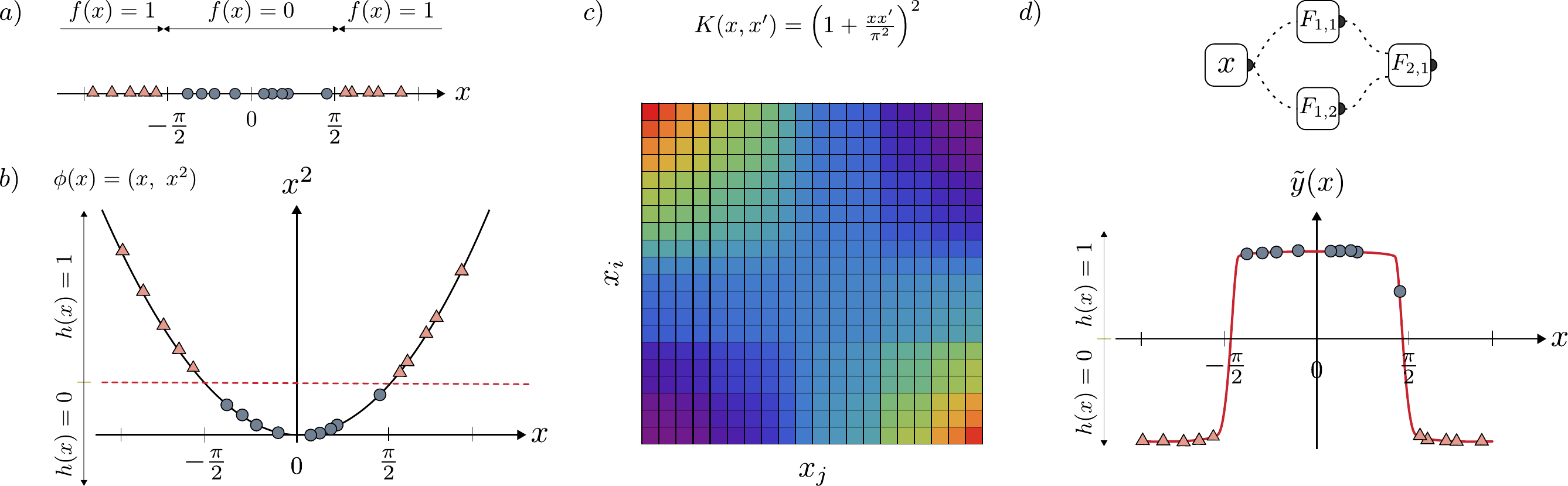}
\caption{Toy-model example and its solution using classical ML techniques. a) We consider a one-dimensional classification problem with input domain $\mathcal{X}=[-\pi,\pi]\subseteq\mathbb{R}$ and labels $\mathcal{Y}=\{0,1\}$ assigned as per the target function $f(x)$ of Eq.~\eqref{eq:toy-model-target}. b) Since there is no linear classifier over the input domain which solves the task, one can map the data onto the high-dimensional space $\mathbb{R}^2$ via the feature map $\phi(x)=(x,\ x^2)$ where a classifying hyperplane does exist (dashed colored line). All points above the classifying hyperplane are assigned a label $h(x)=1$, and all point below a label $h(x)=0$. c) We show the kernel $K(x,x')=\left(1+\frac{xx'}{\pi^2}\right)^2$ for the dataset ordered per increasing value of $x$. This kernel is the backbone of the SVM trained for classification. d) Schematic diagram of a feedforward NN with one   input, two hidden, and one output neuron. After training, the output of $F_{2,1}$ (shown as a solid red line) can be used for classification.     }
\label{chapQML:fig-toy-model-classical}
\end{figure}

As previously mentioned, classical learning algorithms—i.e., choices of hypothesis class $\mathcal{HC}$—come in many shapes and forms. To illustrate this fact, and also to build a bridge to quantum learning algorithms, let us present a toy learning problem as well as several canonical families of classical learning methods. Explicitly, consider a one-dimensional supervised classification problem with input domain $\mathcal{X}=[-\pi,\pi]$ and labels $\mathcal{Y}=\{0,1\}$. Then, let us define the target function
\begin{equation}\label{eq:toy-model-target}
    f(x)=
\begin{cases}
0, & x\in[-\pi/2,\ \pi/2],\\[2pt]
1, & \text{otherwise},
\end{cases}
\end{equation}
so the central interval is class $0$ and the two outer intervals are class $1$ (see Fig.~\ref{chapQML:fig-toy-model-classical}(a)). As per the PAC formalism, one can build a training dataset $\mathcal{S}=\{(x_i,f(x_i))\}_{i=1}^N$ by drawing i.i.d. from the uniform distribution $\mathcal{D}$ on $\mathcal{X}$ (and concomitantly assigning labels via $f$). In the next box we present what is, perhaps, the simplest classifier one can come up with: the linear classifier.

\begin{mybox}{Linear classifiers and feature maps to higher dimensions}

When using a linear classifier, the goal is to find a threshold, draw a hyperplane (or a line in the toy-model example considered) that separates $\mathcal{X}$ into two half-lines, and assign a label to points on each side. However, a linear classifier on $\mathcal{X}$ cannot solve the present task, since no line can be drawn in Fig.~\ref{chapQML:fig-toy-model-classical}(a) such that all points in one class are to the left of the line and all point in the other class to the right. The class with label $1$ is the union of two disjoint outer intervals rather than a single half-line; hence the classes are not linearly separable in the one-dimensional space $\mathcal{X}$.

A standard remedy is to embed the data into a higher-dimensional feature space where a separating hyperplane may exist. For example, consider the feature map $\phi:\mathbb{R}\to\mathbb{R}^2$ and $\phi(x)=(x,\ x^2)$ shown in Fig.~\ref{chapQML:fig-toy-model-classical}(b) and consider linear decision functions of the form
\[
\tilde y(x)=w^\top \phi(x)+b,\qquad \text{ leading to}\qquad  
h(x)=\frac{\operatorname{sign}\!\big(\tilde y(x)\big)+1}{2}\qquad \text{with  $w=(w_1,w_2)$ and } w_1,w_2,b\in\mathbb{R}\,.
\]
Let $z=\phi(x)=(z_1,z_2)=(x,x^2)$. The decision hyperplane in feature space (a line in $\mathbb{R}^2$) is $
w^\top z + b = 0\ \ \Longleftrightarrow\ \ z_2 = -\frac{w_1}{w_2}\,z_1 - \frac{b}{w_2}$ assuming $ w_2\neq 0$ so that $h(x)=1$ for all points above this line, i.e., those with $\tilde y(x)=w^\top z + b \ge 0$ (see Fig.~\ref{chapQML:fig-toy-model-classical}(b)). 

In PAC terms, this corresponds to the hypothesis class of linear separators over the feature map,
\[
\mathcal{HC}_{\mathrm{lin}\circ\phi}=\Bigl\{\,x\mapsto \tfrac{\operatorname{sign}(w^\top \phi(x)+b)+1}{2}\ :\ (w,b)\in\mathbb{R}^2\times\mathbb{R}\Bigr\},
\]
with capacity (and thus generalization) controlled by constraints on $(w,b)$ and sample size. 
\end{mybox}

In practice, explicitly mapping data (which is typically high-dimensional to begin with) into an even higher-dimensional feature space--and operating therein--can be quite computationally expensive. Support vector machines (SVMs) implement this idea implicitly via the kernel trick~\cite{mohri2018foundations}.

\begin{mybox}{Support vector machines and the kernel trick}

Here we recall that a kernel is a function $K:\mathcal{X}\times\mathcal{X}\to\mathbb{R}$ that is symmetric and positive semidefinite. That is, for any set of data points $\{x_i\}_{i=1}^N$, the Gram matrix $[K(x_i,x_j)]_{i,j}$ is positive semi-definite. Equivalently, there exists a (possibly infinite-dimensional) feature map $\phi$ such that $
K(x,x')=\langle \phi(x),\phi(x')\rangle$. Thus inner products in feature space can be computed directly via $K$, without constructing $\phi$. Using, for example, the degree-2 polynomial kernel $K(x,x')=\left(1+\frac{xx'}{\pi^2}\right)^2$ shown in Fig.~\ref{chapQML:fig-toy-model-classical}(c) (which captures quadratic features), the SVM learns the (signed) decision value
\[
\tilde y(x)=\sum_{i=1}^N \alpha_i\, (2 y_i-1)\, K(x_i,x) + b,\qquad
h(x)=\frac{\operatorname{sign}\!\big(\tilde y(x)\big)+1}{2}\,.
\]
The parameters $\alpha_i$ and $b$ are found by minimizing a regularized empirical loss (e.g., the squared loss) subject to margin constraints via a convex dual program (see~\cite{mohri2018foundations} for explicit details).

In PAC terms, the hypothesis class of linear separators is chosen in the feature space induced by $K$
\[
\mathcal{HC}_{\mathrm{ker}}=\Bigl\{\, x\mapsto \tfrac{\operatorname{sign}\!\big(\sum_{i=1}^N \alpha_i (2 y_i-1) K(x_i,x)+b\big)+1}{2}\ :\ (\alpha_1,\ldots,\alpha_N,b)\ \text{satisfy SVM margin/regularization constraints}\Bigr\}.
\]
Now, the capacity (and hence generalization) is controlled by the margin and the norm of the implicit weight vector in feature space, together with the sample size--rather than by explicitly fitting higher-degree decision functions in the raw input.
\end{mybox}

A crucial aspect of the previous methods--irrespective of whether they explicitly or implicitly embed data into higher-dimensional feature spaces--is that they require some form of prior knowledge regarding the data. For instance, one fixes a feature map $\phi$, or a kernel $K$, and then learns a linear separator. In a way, this assumes some intuition about what $\phi$ or $K$ should be, which in turn means that we have some idea of what the defining properties of the dataset are. When such knowledge is not available, we instead need to discover the relevant features from the data itself.

One such approach for learning the correct representation of the data, as well as to combat the curse of dimensionality arising from the fact that $\mathcal{X}$ is usually a very high-dimensional space, is via principal component analysis (PCA)~\cite{mohri2018foundations}. Although PCA does not affect the one-dimensional toy problem as it is aimed at data living in high-dimensional input spaces ($\mathcal{X}=\mathbb{R}^d$ with $d\gg 1$), we still find it convenient to recall it, as its quantum generalization will be studied below in Sec.~\ref{chapQML:subsec:qpca}. PCA is a non-parametric unsupervised learning techniques method that both reduces dimension and extracts the most relevant features: given mean-centered data $X\in\mathbb{R}^{N\times d}$ (rows $\boldsymbol{x}_i^\top-\bar{\boldsymbol{x}}^\top$, with $\bar {\boldsymbol{x}}=\tfrac{1}{N}\sum_i \boldsymbol{x}_i$), it computes the top-$k$ eigenvectors of the empirical covariance $C=\tfrac{1}{N}X^\top X$ (equivalently, the top-$k$ right singular vectors of $X$) and maps inputs via $\boldsymbol{z}=U_k^\top(\boldsymbol{x}-\bar {\boldsymbol{x}})\in\mathbb{R}^k$. This projection maximizes retained variance--by the Eckart–Young theorem--discarding low-variance directions that often correspond to noise or redundancy, and yielding informative, linearly uncorrelated features for downstream learning (e.g., training a kernel SVM on $\boldsymbol{z}_i=U_k^\top(\boldsymbol{x}_i-\bar {\boldsymbol{x}})$).

An alternative approach—learning both the representation and the decision rule end-to-end within a single parametric hypothesis class—is provided by neural networks (NNs)~\cite{anthony2009neural}, which have become some of the most popular classical ML methods in recent years.

\begin{mybox}{Feedforward Neural networks}

In a nutshell, a feedforward NNs is a collection of simple computational units called neurons—functions that apply an affine map followed by a nonlinearity. For instance, a scalar-input neuron takes a real-valued input $x$ and applies the function $\sigma(wx+b)$ with weight $w$, bias $b$, and $\sigma$ the so-called activation function--which sets a threshold for the neuron's output to be non-zero. There are many activation choices, such as the step function $\sigma(x)=1$ if $x\geq x_{\rm threshold}$ and $\sigma(x)=0$ otherwise, in which case the neuron is called a perceptron. Here we illustrate with the sigmoid $\sigma(x)=1/(1+e^{-x})$. but other more modern choices such as ReLU or $\tanh$ would also suffice. As such, consider the small NN of Fig.~\ref{chapQML:fig-toy-model-classical}(d) with one input, two hidden units, and one output. We denote as $F_{1,1}$ and $F_{1,2}$ the hidden neurons and as $F_{2,1}$ the output neuron, respectively, given by 
\[
F_{1,1}(x)=\sigma(w_{1,1}x+b_{1,1}),\qquad
F_{1,2}(x)=\sigma(w_{1,2}x+b_{1,2}),\qquad F_{2,1}(x)=\sigma\big(w_{2,1}\,F_{1,1}(x)+w_{2,2}\,F_{1,2}(x)+b_{2,1}\big)\,.
\]
We train the weights and biases by minimizing an empirical loss (e.g., squared loss) over the data in the training set $\mathcal{S}$, and assign labels as per Fig.~\ref{chapQML:fig-toy-model-classical}(d) using the functions
\[
\tilde y(x):=2\,F_{2,1}(x)-1,\qquad
h(x)=\frac{\operatorname{sign}\!\big(\tilde y(x)\big)+1}{2}\,.
\]

In PAC terms, the hypothesis class is
\[
\mathcal{HC}_{\mathrm{NN}}=\Bigl\{x\mapsto F_{2,1}(x;\theta): \theta\in\Theta\Bigr\},
\]
where $\Theta$ encodes architectural and weight constraints. As such, capacity is governed by architecture/parameter bounds, while the nonlinearity (here, sigmoid) enables representing the nonlinearly separable decision rule. 

An important feature of NNs is that a single hidden layer can realize any linear classifier, making them very flexible models. More generally, a three-layer network (input–hidden–output) can approximate any continuous real-valued function on compact subsets of $\mathbb{R}^k$, with accuracy improving as the number of hidden units increases~\cite{cybenko1989approximation,hornik1991approximation}. While this highlights the large capacity of NNs—and implies that they can, in principle, fit essentially any dataset—the training of the weights and biases leads to highly non-convex optimization landscapes where optimizers may get stuck in poor local minima or linger near saddle points (perhaps as a consequence of the No Free Lunch theorem). Finally, despite the recent heuristic success of deep feedforward architectures with strong inductive biases, their complex structure make it difficult to understand what the learned functions do. This limits our ability to explain their decisions and has motivated concerns about their use as black boxes~\cite{zhang2021survey,antamis2024interpretability}.

\end{mybox}

Beyond NNs and SVMs, there are many other techniques for classification. These include, for instance, the $k$-nearest neighbors (kNN) algorithm, as well as decision trees, but their explanation is beyond the scope of this review.

\subsection{Why learning on quantum computers?}\label{chapQML:subsec:motivation}

In the previous section we have only sketched some of the essentials of learning theory and classical ML. However, these basic concepts suffice to understand how  quantum mechanics can come into play throughout learning tasks. 

Seen through the PAC lens, quantum ingredients can enter at multiple points while keeping the same goal of solving Eq.~\eqref{eq:train-prob}. For instance, the input domain may itself be quantum, such as when $\mathcal{X}\subseteq\mathcal{H}$, where $\mathcal{H}$ denotes the Hilbert space of some quantum system\footnote{When the input domain contains mixed, rather than pure, states, take $\mathcal{X}\subseteq \mathcal{L}(\mathcal{H})=\mathcal{H}\otimes \mathcal{H}^*$, where $\mathcal{L}(\mathcal{H})$ is the space of linear operators acting on $\mathcal{H}$.}~\cite{schuld2019quantum,havlivcek2019supervised}. In this case, data points are quantum states drawn from a distribution over $\mathcal{H}\times\mathcal{Y}$~\cite{sasaki2001quantum,bergou2005universal,cong2019quantum,huang2021provably,perrier2022qdataset,johri2021nearest}. Now, one can consider supervised classification tasks where the training set is
\[
\mathcal{S}=\{\ket{\psi_i}^{\otimes K},y_i\}_{i=1}^N.
\]
Here the tensor-product notation could make it explicit that only $K$ copies of each quantum state are available during training\footnote{This is a subtle but important point. Unlike classical data, quantum states cannot be copied as a consequence of the no-cloning theorem~\cite{wootters1982single}. Because arbitrary repeated access to a given state cannot be assumed, $K$ becomes a key resource parameter. Moreover, since information is obtained via measurements that generally disturb the state, finite $K$ induces an information bottleneck—necessitating judicious use of each copy.}. 

When $\mathcal{X}$ represents classical data, different forms of access lead to different learning models. In the simplest setting, with real vectors ($\mathcal{X}\subseteq\mathbb{R}^d$) or bitstrings ($\mathcal{X}\subseteq\{0,1\}^{\otimes d}$), the “feature map $\to$ linear separator / kernel” recipe can be implemented in a quantum guise by encoding inputs via a unitary $U_\phi\!:x\mapsto \ket{\phi(x)}$ and operating in the induced (potentially exponentially large) Hilbert space~\cite{havlivcek2019supervised,perez2020data,larose2020robust,altares2021automatic,kubler2021inductive,schuld2021effect,ranga2024quantum} (see Sec.~\ref{chapQML:subsec:embedding}). Similarity measures in Hilbert space, such as the standard overlap,
\begin{equation}\label{eq:qkernel-def}
K_q(x,x')=|\langle\phi(x)\mid\phi(x')\rangle|^2\,,
\end{equation}
play the role of quantum kernels~\cite{jerbi2021quantum,schuld2021quantummodels,hubregtsen2021training,shaydulin2021importance,glick2021covariant,gan2023unified,wu2023quantum,henderson2025quantum} and allow training the same SVM decision rule as before, but now in a feature space generated by a quantum circuit (Sec.~\ref{chapQML:subsec:kernel-meth}).

Beyond this, QML offers access models without a classical analogue, such as being given a superposition over the dataset, leading to the quantum example access / quantum PAC setting~\cite{bshouty1995learning} (Sec.~\ref{sec:chapQML:PAC-classical}). Quantum processors can also accelerate subroutines within classical pipelines by implementing certain linear-algebra primitives~\cite{dunjko2018machine} (see Sec~\ref{sec:chapQML:qml-linear-algebra}), e.g., quantum PCA~\cite{rebentrost2014quantum} for dimensionality reduction, or HHL-style solvers for linear systems~\cite{harrow2009quantum}--under specific data-access assumptions-- or even implement generalizations of classical techniques such as kNN algorithm~\cite{schuld2014quantum,wisniewska2018recognizing,ruan2017quantum,zardini2024quantum}, as well as decision trees~\cite{farhi1998quantum,lu2014quantum,kumar2025q}.

Finally, QML broadens what is meant by “learning.” For instance, one can learn states (e.g., given copies, predict outcomes for all—or a restricted family of—measurements) or learn processes (e.g., given an unknown unitaries or, more generally, quantum channels, have the ability to  implemented them or make predictions from their on future). In PAC terms, the state or the process becomes the unknown concept to be learned (see Sec.~\ref{sec:chapQML:PAC-quantum}).

As one can rapidly see, there are many ways in which the laws of quantum mechanics can interplay with the mathematical framework of computational learning theory. The real question is whether there is a genuine benefit to using quantum computers\footnote{For the foreseeable future quantum computers are likely to remain expensive, specialized resources, so their use must therefore be properly justified.}. That is, whether there is any advantage over purely classical means, be it reduced runtime, different sample/complexity tradeoffs (including the copy complexity $K$ in quantum settings), or enhanced representational power. This question guides the remainder of the review.

\section{Variational quantum machine learning}

\begin{figure}[t]
\centering
\includegraphics[width=.9\textwidth]{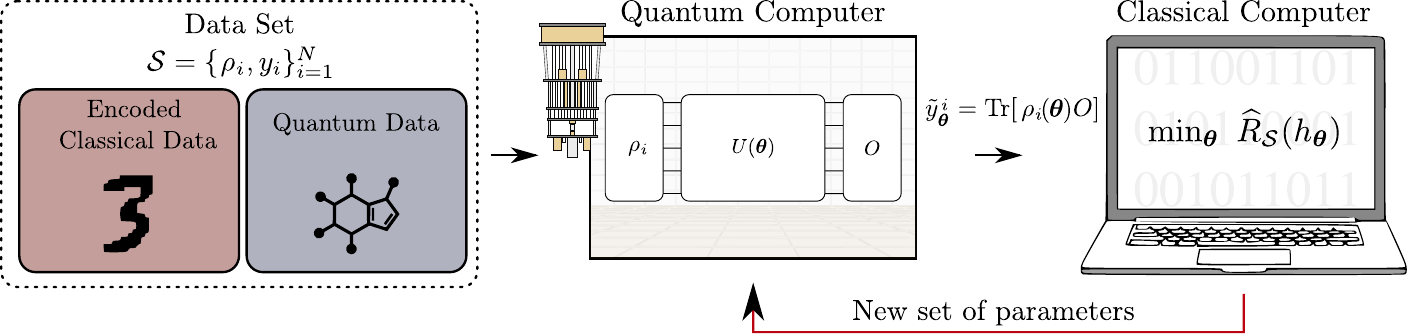}
\caption{Schematic representation of an archetypal variational QML pipeline.  States from a supervised learning dataset $\{\rho_i,y_i\}_{i=1}^N$ are sent through a parametrized quantum circuit (or quantum neural network) $U(\boldsymbol{\theta})$, and one can estimate the quantities $\tilde{y}_{\boldsymbol{\theta}}^i={\rm Tr}[\rho_i(\boldsymbol{\theta})O]$ for $\rho_i(\boldsymbol{\theta})=U(\boldsymbol{\theta})\rho_iU^\dagger(\boldsymbol{\theta})$. These expectation values are fed to a classical optimizer that computes an empirical loss $\widehat{R}_{\mathcal{S}}(h_{\boldsymbol{\theta}})$ (where $h_{\boldsymbol{\theta}}$ denotes the output parametrized hypothesis), and leverages classical optimizers to find the new set of parameters which solves the optimization problem $\min_{\boldsymbol{\theta}} \ \widehat{R}_{\mathcal{S}}(h_{\boldsymbol{\theta}})$. }
\label{chapQML:figvarQML}
\end{figure}

Variational QML is nowadays the most popular and widely used approach to learning  from classical and from quantum data. In a nutshell, this paradigm adopts the parametric, data-driven view of modern ML, where one solves a learning task by encoding it as an optimization problem, chooses a flexible family of models, and lets the data select—via training—the specific hypothesis that solves the task. In the language of PAC, this means the hypothesis class $\mathcal{HC}_{\boldsymbol{\theta}}$ is parametrized and the parameters $\boldsymbol{\theta}$ are trained by minimizing an empirical loss that quantifies task performance. Most variational QML models follow the hybrid computational paradigm of Fig.~\ref{chapQML:figvarQML}: a quantum device prepares states of interest (which either encode classical information or are themselves quantum data), applies a parametrized evolution that coherently processes the state, and performs a measurement at the output. Outputs from the quantum device typically arrive as finite-shot estimates of expectation values,
\begin{equation}\label{eq:cost}
    \tilde y_{\boldsymbol\theta}(x)=\mathrm{Tr}[\rho(x,\boldsymbol{\theta})\,O]\,,
\end{equation}
where $\rho(x,\boldsymbol{\theta})$ is the data encoded state, $\boldsymbol{\theta}\in\Theta$ are trainable parameters in a domain $\Theta$, and $O$ is a task-dependent observable. The estimates $\tilde y_{\boldsymbol\theta}(x)$ are then fed to a classical computer to assemble an empirical loss $\widehat{R}_{\mathcal{S}}(h)$ and to choose the next parameters or measurements, thereby tackling the optimization problem in Eq.~\eqref{eq:train-prob} with $h$ induced by $\boldsymbol{\theta}$.

From the previous one can see the appeal of variational QML. This paradigm is versatile and appears conceptually simple: once a faithful loss is specified, much of the heavy lifting can be outsourced to a classical optimizer. Moreover, not having to execute the full computation exclusively on a quantum device can reduce the burden on quantum hardware. As such, variational QML often deploys learning algorithms that require fewer gates or ancillas, making them better suited for noisy intermediate-scale quantum (NISQ)~\cite{preskill2018quantum} processors.

In the next sections, we will describe some of the basic ingredients of variational QML (data embedding, quantum kernel methods, and quantum neural networks). For a more detailed review, we refer the reader to Refs.~\cite{cerezo2020variationalreview,bharti2021noisy,gujju2024quantum,wang2024comprehensive}.

\subsection{Embedding classical data\label{chapQML:subsec:embedding}}
When learning from classical data, the first step in a variational QML setting is to embed it into the quantum Hilbert space $\mathcal{H}$~\cite{lloyd2020quantum}. A common choice for bitstrings, i.e., when $\mathcal{X}=\{0,1\}^d$, is binary-basis embedding: given $x\in\mathcal{X}$, prepare the computational-basis state $\ket{x}$. This approach is natural for binary features, and downstream circuits can act directly on the computational basis. When the input consists of real-valued vectors ($\mathcal{X}=\mathbb{R}^d$), a widely used option is amplitude encoding~\cite{maheshwari2021variational}. Now, given a normalized $x\in\mathbb{R}^{d}$, prepare the state
\[
\ket{x}\;=\;\sum_{i=0}^{d-1} x_i \ket{i}.
\]
This encoding maximizes feature density per qubit but generic state preparation can be quite demanding, as  determining a circuit that prepares $\ket{x}$ requires, in the worst case, circuit depth scaling linearly with the Hilbert space dimension  ${\rm dim}(\mathcal{H})=d$, potentially making this encoding impractical. When dealing with $n$-qubit quantum computers, where $d=2^n$, a simpler alternative that admits shallow circuits is angle-rotation embedding. Here, where real-valued features are used as angles for gates that are applied to subsets of qubits. For instance, on qubit $j$ apply a general rotation $R_{\boldsymbol{v}_j}(x_j)=e^{-i x_j \boldsymbol{v}_j\cdot\boldsymbol{\sigma}/2}$ with $\boldsymbol{\sigma}_j=(X_j,Y_j,Z_j)$ and $\boldsymbol{v_j}\in\mathbb{R}^3$, possibly followed by fixed entangling gates. A simple instance is
\[
V_{\boldsymbol{v}}(x)\;=\;\bigotimes_{j=1}^n R_y(x_j)\; = \; \bigotimes_{j=1}^n e^{-i \nu_j x_j    Y_j },
\]
while more expressive feature maps include products of commuting, data-dependent exponentials such as
\[
V_{\boldsymbol{v}}(x)\;=e^{i\sum_j v_j x_j Z_j\;+\;i\sum_{j<k}v_{jk}\, x_j x_k\, Z_j Z_k},
\]
where the coefficients $v_j$ and $v_{jk}$ may themselves be trainable~\cite{havlivcek2019supervised,mitarai2018quantum}.

Once the classical data is mapped onto a quantum computer, one can already explore whether there is a linear classifier over $\mathcal{H}$. For instance, we can showcase this fact for the toy-model example of Sec.~\ref{chapQML:subsec:toy} by encoding the data onto a single qubit.

\begin{figure}[t]
\centering
\includegraphics[width=1\textwidth]{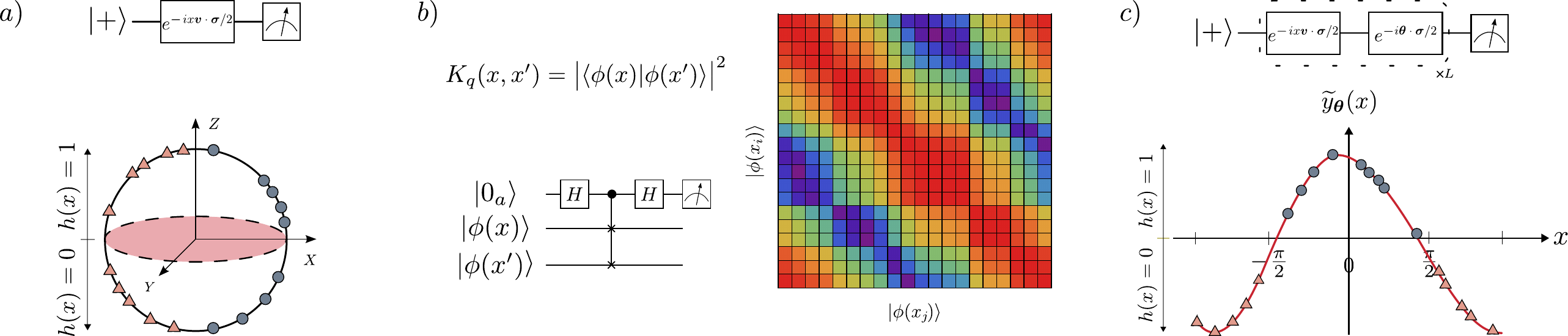}
\caption{Toy-model example and its solution using QML techniques on a single qubit. a) A feature map encodes data as gate-rotation angles, embedding points into the qubit Hilbert space $\mathcal{H}=\mathbb{C}^2$ (a higher-dimensional feature space relative to the 1D input). A classifying hyperplane is then trained by measuring a rotated Pauli observable with trainable parameters. We show the encoded data for $\boldsymbol{v}=(0,1,0)$. b) One can also leverage the encoded data to estimate—via the SWAP-test circuit—the quantum kernel $K_q(x,x')=\big|\langle\phi(x)\vert\phi(x')\rangle\big|^2$. c) A QNN is implemented as a sequence of data encoding unitaries and parametrized trainable gates. After training, the expectation value of $
    \tilde y_{\boldsymbol\theta}(x)=\mathrm{Tr}[\rho(x,\boldsymbol{\theta})\,O]$ for $O=Z$ (shown as a solid red line) can be used for classification.}
\label{chapQML:fig-toy-model-quantum}
\end{figure}

\begin{mybox}{Linear classifiers via feature maps to higher dimensions.}

Since the data of the toy model in Eq.~\eqref{eq:toy-model-target} is not linearly separable in $\mathcal{X}$, each point is embedded into the qubit Hilbert space $\mathcal{H}=\mathbb{C}^2$ via the rotation-angle feature map $V_\phi$
\[
\ket{\phi(x)}=e^{-i x\,\boldsymbol{v}\cdot \boldsymbol{\sigma}/2}\ket{+}\,,
\]
where $\boldsymbol{\sigma}=(X,Y,Z)$ and $\boldsymbol{v}=(v_1,v_2,v_3)\in\mathbb{R}^3$ controls the encoding axis (with $\ket{+}=(\ket{0}+\ket{1})/\sqrt{2}$). This maps $x\in[-\pi,\pi]$ to points on the Bloch sphere, i.e., a nonlinear embedding into a higher-dimensional geometric space. In Fig.~\ref{chapQML:fig-toy-model-quantum}(a) we show the case $\boldsymbol{v}=(0,1,0)$.

One can then define a linear decision function in feature space by measuring the Pauli observable $Z$, 
\[
\tilde y_{\boldsymbol{v}}(x)={\rm Tr}\!\left[\ket{\phi(x)}\bra{\phi(x)}\,Z\right],\qquad
h(x)=\frac{\operatorname{sign}\!\big(\tilde y_{\boldsymbol{v}}(x)\big)+1}{2}\,.
\]
By tuning $\boldsymbol{v}$ one aligns the great-circle separator with the target labeling.

This induces the hypothesis class
\[
\mathcal{HC}_{\mathrm{lin}\circ\phi}^{(q)}=\Bigl\{\,x\mapsto \frac{\operatorname{sign}\!\big(\tilde y_{\boldsymbol{v}}(x)\big)+1}{2}\,:\ \boldsymbol{v}\in\mathbb{R}^3\Bigr\},
\]
and training amounts to empirical risk minimization over $\boldsymbol{v}$. In practice we cannot estimate $\tilde y_{\boldsymbol{v}}(x)$ to arbitrary precision, but rather via a finite set of $N_s$ shots. Measuring in the computational basis yields outcomes $+1$ (for $\ket{0}$) and $-1$ (for $\ket{1}$), and a frequentist estimator for $\tilde y_{\boldsymbol{v}}(x)$ is $\frac{N_0-N_1}{N_s}$, where $N_0$ and $N_1$ are the counts of outcomes $0$ and $1$, respectively. This estimator is unbiased with variance $\frac{1-\tilde y_{\boldsymbol{v}}(x)^2}{N_s}$ (for $\pm1$ outcomes), so the standard error due to shot noise scales as $1/\sqrt{N_s}$. These shot-based estimates of $\tilde y_{\boldsymbol{v}}(x)$ supply the labels to the optimizer, while generalization is assessed as in Eqs.~\eqref{eq:emp-loss}–\eqref{eq:pop-risk}.

\end{mybox}

While angle-based encodings such as the one above are popular, care is needed when using them as  there is no universally “right’’ way of placing classical data into $\mathcal{H}$, and the choice is often heuristic and task dependent. For example, in the setting above, not all values of $\boldsymbol{v}$ lead to accurate classification when measuring $Z$. More broadly, the quantum embedding can introduce periodicity and scale sensitivity (e.g., angle wrap-around and feature normalization), interact non-trivially with entangling patterns--leading to state that are mapped to randomly parts of Hilbert space--and substantially affect performance~\cite{schuld2019quantum,schuld2021effect,ranga2024quantum,thanaslip2021subtleties}. Indeed, embedding design is still an active area of research.

\subsection{Quantum kernel methods}\label{chapQML:subsec:kernel-meth}

Having access to data (classical or quantum) in a Hilbert space enables kernel methods and classification via SVMs. For this purpose, one needs to define a function $K:\mathcal{X}\times\mathcal{X}\to\mathbb{R}$ that is symmetric and positive semidefinite. For instance, a basic choice is the Hilbert–Schmidt kernel
\[
K_{ij}\;=\;\mathrm{Tr}\big[\rho_i\,\rho_j\big],
\]
where we have used the notation  $\rho_i$ instead of $\rho(x_i)$ to account for the fact that $\rho_i$ can be taken from a quantum dataset. More generally, other kernels have been proposed based on different ways to compare quantum states, such as the projected quantum kernel that uses the inner-product between the  marginals--i.e., reduced states--of $\rho_i$ and $\rho_j$-- and thus checks how similar the data point are  locally~\cite{schuld2019quantum, havlivcek2019supervised,peters2021machine,sancho2022quantum,Blank2020Quantum,rodriguez2025neural,canatar2022bandwidth,huang2021power}. 

Once the kernel is defined, one trains standard learners  (typically an SVM) on the Gram matrix that was estimated on a quantum device. Since the learner itself is classical, any genuine advantage must come either from the embedding (it exposes nontrivial features of $\mathcal{X}$ that classical kernels would not capture with comparable resources) or from the fact that estimating the kernel entries cannot be efficiently simulated classically (i.e., the kernel trick is not available)~\cite{liu2021rigorous,wang2021towards,jager2023universal,incudini2024toward}. In other words, for classical data the SVM’s performance is driven by the encoding: the kernel inherits the inductive bias of the feature map.

\begin{mybox}{Quantum kernels and support vector machines}

After embedding the data on the Bloch sphere, one can use the quantum kernel
\[
K_q(x,x')=\big|\bra{\phi(x)}{\phi(x')}\rangle\big|^2,
\]
which can be estimated with the SWAP test~\cite{buhrman2001quantum,cincio2018learning} shown in Fig.~\ref{chapQML:fig-toy-model-quantum}(b). Therein,  we denote the ancilla qubit with the sub-index $a$, and the two system registers hold $\ket{\phi(x)}$ and $\ket{\phi(x')}$. The circuit shown applies the following sequence of transformations
\begin{align*}
\ket{0}_a\ket{\phi(x)}\ket{\phi(x')} &\xrightarrow{H_a} 
\frac{1}{\sqrt{2}}\!\left(\ket{0}_a+\ket{1}_a\right)\ket{\phi(x)}\ket{\phi(x')}
\xrightarrow{\,\mathrm{c\text{-}SWAP}\,}
\frac{1}{\sqrt{2}}\!\left(\ket{0}_a\ket{\phi(x)}\ket{\phi(x')}+\ket{1}_a\ket{\phi(x')}\ket{\phi(x)}\right)\\
&\xrightarrow{H_a}
\frac{1}{2}\!\left[\ket{0}_a\big(\ket{\phi(x)}\ket{\phi(x')}+\ket{\phi(x')}\ket{\phi(x)}\big)
+\ket{1}_a\big(\ket{\phi(x)}\ket{\phi(x')}-\ket{\phi(x')}\ket{\phi(x)}\big)\right]\,.
\end{align*}
Above, $\mathrm{c\text{-}SWAP}$ denotes the controlled SWAP operation. 
A direct calculation reveals that the probability of measuring the qubit on the zero state is $\Pr[a{=}0]=\tfrac{1+|\braket{\phi(x)}{\phi(x')}|^2}{2}$ and hence 
\[
\langle Z_a\rangle=\Pr[a{=}0]-\Pr[a{=}1]=\big|\langle\phi(x)\ket{\phi(x')}\big|^2=K_q(x,x').
\]
When estimating this expectation value with $N_s$ shots, the frequentist estimator of the kernel is $\frac{N_0-N_1}{N_s}$ which again comes accompanied by a standard error due to shot noise scaling as $1/\sqrt{N_s}$.

Using SVMs with $K_q$ simply replaces the kernel in the hypothesis class. The training proceeds as before with an empirically estimated Gram matrix whose entries are shot-noisy estimates $\widehat{K}_q(x_i,x_j)$.

\end{mybox}

 When using quantum kernels there are several import facts one must consider. First, bandwidth and global rescaling of the encoding strongly affect inductive bias: too large leads to underfitting; too small to overfitting~\cite{kubler2021inductive,canatar2022bandwidth,huang2021power,gil2024expressivity}. While proper tuning can mitigate the decay of kernel values with qubit count, ill defined feature maps and kernels can lead to  diagonal kernels, and thus to poor generalization. Second, quantum-information properties of the embedding and data structure play a key role in its model's performance. For instance, fidelity kernels on ``far'' states make the Gram matrix nearly diagonal (little shared structure), while projected/local kernels on highly entangled states can become uninformative because reduced states approach maximally mixed, pushing kernel entries toward a constant. These effects do not rule out quantum kernels, but they reinforce the point that the encoding—and its induced inductive bias—ultimately sets performance.

\subsection{Parametrized quantum circuits and quantum neural networks}

One of the central requirements in variational QML is to choose a tractable way to parametrize quantum models. Unlike classical ML--where linear-algebra toolkits abound--QML must respect the postulates of quantum mechanics: coherent operations are linear/unitary and measurements collapse the information within the state. A natural parametrization comes from controlled time evolutions. For a Hamiltonian $H$, the state evolves as $\ket{\psi(t)}=U(t)\ket{\psi(0)}$ with $U(t)=e^{iHt}$, so treating $t$ as a tunable parameter gives a one-parameter gate. More flexibly, switching between implementable Hamiltonians $\mathcal{G}=\{H_1,\dots,H_p\}$ for times $\theta_1,\dots,\theta_p$ yields the standard parametrized quantum circuit (PQC), or quantum neural network (QNN), ansatz 
\begin{equation}\label{eq:PQC}
    U(\boldsymbol{\theta})=\prod_{l=1}^LU_l(\theta_l)\,,\quad \text{with}\quad U_l(\theta_l)=e^{-i \theta_l H_l}\quad \text{and}\quad H_l\in \mathcal{G}\,.
\end{equation}
The specific choice of the set of generators $\mathcal{G}$ depends on the task at hand. For instance, one could simply select the Hamiltonians that are native to the quantum device, leading to the so-called hardware efficient ansatz~\cite{kandala2017hardware}. Typically here, entangling gates follow the device connectivity to avoid unnecessary compiling overheads~\cite{khatri2019quantum,he2021variational,moro2021quantum} and are implemented on a brick-like fashion on alternating pairs of qubits or follow a ladder-like structure. Other choices include the use of restricted gate sets (such as matchgates~\cite{cherrat2023quantum,jozsa2008matchgates,wan2022matchgate,de2013power,oszmaniec2022fermion, matos2022characterization, diaz2023parallel}) or unitaries that are inspired by the task at hand and can encode information about the problem one is trying to solve~\cite{wecker2015progress,wiersema2020exploring,larocca2021diagnosing,park2023hamiltonian}. In this case, one recovers the unitary coupled singles and doubles (UCCSD) ansatz~\cite{peruzzo2014variational,taube2006new,lee2018generalized,arrazola2022universal} for quantum chemistry problems, the quantum alternating operator ansatz (QAOA) used in combinatorial optimization~\cite{farhi2014quantum,hadfield2019quantum,bartschi2020grover,fuchs2022constrained,tate2021classically}, or equivariant circuits that are constructed to respect the dataset's symmetries and only explore symmetry-respecting solution spaces~\cite{lee2021towards,cherrat2023quantum,larocca2021diagnosing,monbroussou2023trainability,larocca2022group,ragone2022representation,nguyen2022atheory,meyer2022exploiting,skolik2022equivariant,volkoff2021large,sauvage2022building,schatzki2022theoretical,kazi2023universality,kerenidis2021classical,maccormack2020branching,Chang2023Approximately,dong2024Z2,west2023reflection}. In addition, quantum generalizations of archetypal classical NNs have proposed such as the quantum convolutional neural network~\cite{cong2019quantum,bermejo2024quantum,liu2023model,zapletal2024error,pesah2020absence,henderson2020quanvolutional}, recurrent quantum neural network~\cite{bausch2020recurrent,gonon2025feedback,siemaszko2022rapid,Takaki2021Learning}, quantum graph neural networks~\cite{verdon2019quantumgraph,forestano2024comparison} and quantum feedforward neural networks~\cite{wan2017quantum}, among many others. Finally, while Eq.~\eqref{eq:PQC} assumes a fixed structure in the PQC, one can also use  strategies which adaptively change the structure of the  ansatz~\cite{tang2019qubit,grimsley2019adaptive,grimsley2022adapt,anastasiou2024tetris,cincio2018learning, bilkis2021semi,du2020quantum,sim2021adaptive,zhang2021mutual,tkachenko2020correlation,claudino2020benchmarking,rattew2019domain,chivilikhin2020mog,zhang2020differentiable,wada2022sequential}.

In the box below, we complete the toy-model analysis by showing how this learning task can be solved with PQCs.

\begin{mybox}{Parameterized quantum circuits and quantum neural networks}

Let us begin by assuming that the classical data is encoded onto a quantum state as in the linear classifier box. That is, $\ket{\phi(x)}=e^{-i x\,\boldsymbol{v}\cdot \boldsymbol{\sigma}/2}\ket{+}$ (see Sec.~\ref{chapQML:subsec:embedding}). It is not hard to see that sending such a state through a general single qubit PQC of the form $e^{-i\,\boldsymbol{\theta}_l\cdot \boldsymbol{\sigma}/2}$ is equivalent to simply changing the vector $\boldsymbol{v}$ to a different one as $\boldsymbol{v}'(\boldsymbol{\theta})$. Hence, this  simply recovers the linear classifier result from above (this result showcases the important fact that in the quantum realm PQC-based QML methods are ultimately kernel techniques~\cite{schuld2021quantum}). To sidestep this issue and increase the model expressiveness a la classical NN, we can instead use the single-qubit “data reuploading’’~\cite{perez2020data} architecture shown in Fig.~\ref{chapQML:fig-toy-model-quantum}(c) which alternates data encoders and trainable rotations
\[
W(x,\boldsymbol{v},\boldsymbol{\theta})=\prod_{l=1}^{L}\!\left[e^{-i\,\boldsymbol{\theta}_l\cdot \boldsymbol{\sigma}/2}\,e^{-i\,x\,\boldsymbol{v}_l\cdot \boldsymbol{\sigma}/2}\right],\qquad
\boldsymbol{\theta}=\{\boldsymbol{\theta}_l\}_{l=1}^L,\ \boldsymbol{v}=\{\boldsymbol{v}_l\}_{l=1}^L,\ \boldsymbol{\theta}_l,\boldsymbol{v}_l\in\mathbb{R}^3,
\]
(where the rightmost factor acts first). Applying this unitary to the input state $\ket{+}$ yields the parametrized state
\[
\ket{\phi(x;\boldsymbol{v},\boldsymbol{\theta})}=W(x,\boldsymbol{v},\boldsymbol{\theta})\ket{+}.
\]
At the output of the circuit we measure $Z$ and define
\[
\tilde y(x):=2\,\bra{\phi(x;\boldsymbol{v},\boldsymbol{\theta})}Z\ket{\phi(x;\boldsymbol{v},\boldsymbol{\theta})}-1,\qquad
h(x)=\frac{\operatorname{sign}\!\big(\tilde y(x)\big)+1}{2}.
\]
One can train $\{\boldsymbol{v}_l,\boldsymbol{\theta}_l\}$ by minimizing an empirical loss (e.g., squared or logistic) over $\mathcal{S}$. The depth $L$ controls the set of realizable functions $x\mapsto \tilde y(x)$. In particular, data reuploading achieves universal approximation on one qubit (i.e., it can approximate a broad class of functions arbitrarily well)~\cite{perez2020data}. However, training of circuits with large $L$ can be quite hard, and the optimizer might get stuck in sub-optimal local minima (perhaps again, as a consequence of the No Free Lunch theorem).

This fixes a hypothesis class $\mathcal{HC}_{\mathrm{QNN}}=\{x\mapsto h(x;\boldsymbol{v},\boldsymbol{\theta})\}$ over which one performs empirical risk minimization. As before, expectation values are estimated with $N_s$ shots, introducing a shot-noise standard error that scales as $1/\sqrt{N_s}$.

\end{mybox}

With the parametrization in place, one can analyze a model’s expressiveness and capacity~\cite{larocca2021diagnosing,zeier2011symmetry,wiersema2023classification,wierichs2023symmetric,aguilar2024full,abbas2020power,haug2021capacity,bu2021onthestatistical,bu2021effects,bu2021rademacher,gyurik2021structural}, as well as its generalization capabilities~\cite{caro2021generalization,caro2022outofdistribution,caro2020pseudo,popescu2021learning,banchi2021generalization,gibbs2022dynamical,manzano2025approximation}. For instance, when using a data reuploading scheme where layers of angle encoding are followed by PQCs  the  expectation values such
as those in Eq.~\eqref{eq:cost} take the form 
\[\tilde y_{\boldsymbol\theta}(x)={\rm Tr}\left[\left(\prod_{l=1}^LU_l(\boldsymbol{\theta_l})V_l(x)\right)\ket{\psi_0}\bra{\psi_0}\left(\prod_{l=1}^L V_l^\dagger(x)U_l^\dagger(\boldsymbol{\theta_l})\right) O\right]\,,\]
for some fiducial state $\ket{\psi_0}$. Then, under relatively mild-assumption on the data encoding unitaries the output will admit a Fourier-like decomposition of the form~\cite{schuld2021effect,gil2020input}
\begin{equation}\label{eq:Fourier}
    \tilde y_{\boldsymbol\theta}(x)=\sum_{\omega\in\Omega}c_\omega(\boldsymbol{\theta})e^{\langle\omega,x\rangle}\,,
\end{equation}
where the frequencies $\omega$ and their domain $\Omega$ depends only on the gates in $V(x)$. Equation~\eqref{eq:Fourier} characterizes the type of function over $x$ that the model can represent. In particular, the complexity of the model can be increased by the so-called data reuploading techniques~\cite{perez2020data,rodriguez2024training} where encoding layers and alternated with PQC layers (see the box above). 

More generally, PAC/VC tools can bound generalization on PQC-based variational QML models. If a PQC has $L$ parametrized local unitaries (each acting on $\mathcal{O}(1)$ qubits) and observables are bounded, then with high probability over $\mathcal{S}$ of size $N$ the generalization gap satisfies
\begin{equation}\label{eq:gap}
    \mathrm{gap}_{\mathcal{S}}(h)\in\mathcal{O}\left(\sqrt{\frac{L\log(L)}{N}}\right)\,.
\end{equation}
Equivalently, for any $\varepsilon>0$, with high confidence, one has $\mathrm{gap}_{\mathcal{S}}(h)\le \varepsilon$ whenever $N \gtrsim \frac{L\log (L)}{\varepsilon^2}$. 
Thus the required number of training examples scales effectively linearly with the number of trainable parameters.

Finally, a note on terminology. Although “PQC’’ and “QNN’’  are used interchangeably in the literature, a QNN is fundamentally different from a classical NN. Classical NNs hinge on explicit nonlinear activations and unrestricted fan-out/copying of intermediate representations. Neither exists natively in closed-system quantum mechanics: evolution is linear/unitary and the no-cloning theorem forbids copying unknown quantum states~\cite{wootters1982single}. Any effective nonlinearity arises via measurement and classical post-processing (or measurement-based feedback), changing the computational model. This makes apples-to-apples comparisons with classical NNs unreliable, and simple metrics such as the number of neurons or layers are not meaningful proxies for capability in PQCs/QNNs.

\subsection{Challenges and limitations for variational QML}

While variational QML is perhaps the most studied approach to QML, this computational paradigm has several shortcoming that one must understand ~\cite{cerezo2022challenges}. Perhaps the main limitation for variational QML is that it generally has no theoretical guarantees. That is, when choosing a models (e.g., a given hypothesis class $\mathcal{HC}_{\boldsymbol{\theta}}$) there is no assurance that a solution even exists within. In addition, even if it exists, it is not necessarily true  that a classical optimizer will be able to solve Eq.~\eqref{eq:train-prob}. This means that variational QML is mostly a heuristic field, oftentimes guided by good practices and rules of thumb. While a priori one could argue that this is not an issue, as classical ML is also a heuristic field where models regularly outperform their expected theoretical capabilities, QML heuristics  have been reported in which the exact same model can exhibit widely different performance on the same task. This underscores the critical role of hyperparameter fine-tuning~\cite{bowles2024better,qian2021dilemma}, and the dangers of relying on single, non-repeated experiments that are oftentimes not validated on other works. Moreover, the situation becomes more complex when one takes into account the fact that the ultimate arbiter--large-scale quantum computers with sufficiently low noise levels--is not yet available, limiting tests at scale and confining our understanding to narrow settings where theoretical proofs are available or restricted simulations and benchmarks scenarios (often at the small-scale) with varying degrees of reliability.

To make the situation become complicated, there exists a large body of literature analytically studying the properties of the optimization landscape that one must navigate to solve Eq.~\eqref{eq:train-prob}~\cite{anschuetz2024unified}. Within this context, the most studied phenomenon is that of barren plateaus~\cite{larocca2024review,mcclean2018barren,cerezo2020cost,ragone2023unified,wang2020noise,franca2020limitations}, whereby the differences between points in the loss function landscape vanish exponentially with the system size (or the number of qubits, $n$). Given a loss as un Eq.~\eqref{eq:cost}, then such phenomenon could exist on average over the landscape, i.e.,  $\mathbb{E}_{\boldsymbol{\theta}_1,\boldsymbol{\theta}_2\sim\Theta}[\tilde y_{\boldsymbol\theta_1}(x)-\tilde y_{\boldsymbol\theta_2}(x)]\in\mathcal{O}(1/{\rm dim}(\mathcal{H}))$, or for all points, i.e., $\forall \boldsymbol{\theta}_1,\boldsymbol{\theta}_2\in\Theta$ one has $(\tilde y_{\boldsymbol\theta_1}(x)-\tilde y_{\boldsymbol\theta_2}(x))\in\mathcal{O}(1/{\rm dim}(\mathcal{H}))$. Given that the uncertainty in expectation values computed with $N_s$ shots scale as $1\sqrt{N_s}$ shots, in a barren plateau an exponential number of measurements are needed to distinguish small differences between points in the landscape and find an optimizing direction. Thus precluding the train models at the large scale. Ultimately, barren plateaus can be understood as a form of curse of dimensionality arising from the fact that quantities such as $\tilde y_{\boldsymbol\theta}(x)$ in Eq.~\eqref{eq:cost} compare objects in exponentially large spaces (see~\cite{larocca2024review} for an extensive review on this topic). While some techniques have been proposed to mitigate this curse of dimensionality, it actually turns out that most techniques for preventing barren plateaus actually dequantize the model and make it classically simulable\footnote{Intuitively, if a barren plateaus exists because one is trying to manipulate data and  in a very large dimensional vector space, one avoids this concentration phenomenon by restricting a smaller subspace. However, such restriction makes the whole dynamics of the PQC classically simulable, as one needs to only simulate the smaller subspace of interest, rather than the whole Hilbert space~\cite{cerezo2023does}.}~\cite{bermejo2024quantum,cerezo2023does,angrisani2024classically,lerch2024efficient,anschuetz2024arbitrary,mele2024noise,shin2024dequantising,ermakov2024unified,miller2025simulation,rudolph2022synergy,fontana2023classical} (e.g., one could learn the low frequencies of Eq.~\eqref{eq:Fourier} to simulate the loss~\cite{shin2024dequantising, gil2024relation,landman2022classically,sweke2023potential,sahebi2025dequantization,o2025efficient,shaffer2023surrogate,gustafson2024surrogate}). More than barren plateaus, it has also been shown that variational QML optimization landscapes are plagued with sub-optimal local minima~\cite{anschuetz2021critical,bittel2021training,larocca2021diagnosing,fontana2022nontrivial,anschuetz2022beyond} where the optimizer can get trapped. Furthermore, no practical approach is known to avoid local-minima issues on standard PQCs. For instance, to mitigate local minima, one requires a circuit with a number of parameters that typically scales with the Hilbert space dimension, which in turn implies from Eq.~\eqref{eq:gap} that the model will only perform well  if it is trained on a number of training points scaling as $\Omega({\rm dim}(\mathcal{H})\log({\rm dim}(\mathcal{H})))$~\cite{larocca2021theory,garcia2023effects,kiani2020learning,campos2021abrupt,you2022convergence}. 

It is also worth mentioning that when training PQCs with gradient methods one must estimate $\partial \tilde y_{\boldsymbol\theta}(x)/\partial \theta_\ell$. For many common costs and generators, a technique known as the parameter-shift rule does this with the same circuit used for $\tilde y_{\boldsymbol\theta}(x)$, just with simple angle shifts~\cite{mitarai2018quantum,schuld2019evaluating,markovich2024parameter}. The catch is that each gradient entry has to be measured separately. Hence, with $T$ parameters, computing the full gradient costs $\mathcal{O}(T)$ circuit evaluations (times the shots per estimate), which quickly becomes expensive for large $T$. Moreover, unlike classical NNs, genuine quantum backpropagation is ruled out by basic quantum constraints, yielding no-go results for standard backpropagation tricks~\cite{abbas2023quantum}. Finally, we note that standard parameter update rules require the implementation of small rotation angles (this is also true  during data encoding). While this is not an issue for NISQ devices, their deployment on fault-tolerant quantum computers can be prohibitively large, as small angle rotations come at a very high $T$-gate cost~\cite{koczor2024probabilistic,ross2014optimal}. The degree to which variational QML can be adapted to fault-tolerant architectures is still a topic of study.

\section{Quantum computational learning theory and some provable separations}
\label{chapQML:sec:learning_theory}

The previous sections have laid the groundwork for what it means to learn from data within the PAC and VC theory, as well as introduced basic ideas for how quantum and learning can be combined within the data-driven parametric variational QML framework. However, we have also discussion the (many) limitations of variational QML and the fact that it is a heuristic field with little-to-no theoretical guarantees. This can prompt the reader to ask: ``\textit{Are there tasks where a quantum learner can actually solve a task more efficiently (under some metric) than any classical learners?}'' In this section we will give a positive answer to this question. To do so, we will switch gears away from parametric models, and instead consider non-parametric QML: We started from a well-posed task, designed a clever algorithm tailored to that task, and estimate how many resources where used.

More specifically, we will introduce the key concepts behind quantum PAC (we also refer the reader to~\cite{dunjko2018machine,arunachalam2017guest}), and present scenarios where provable separations between classical and quantum learners exist. We will first focus on the task of learning classical concepts from quantum examples, and then move onto the task of learning quantum concepts. As we will see, learning separations come in many different shapes and forms which make sense within their particular setting~\cite{gyurik2022establishing}. However, there is no single definition that we will focus on, and instead opt to show how by tweaking subtle differences in definitions one can prove that a model is formally more efficient than another.

\subsection{Quantum PAC and beyond, learning classical concepts} \label{sec:chapQML:PAC-classical}

Quantum PAC is a computational learning theory which formally extends standard PAC by allowing quantum resources, such as quantum computation or access to quantum examples (superposition of examples)~\cite{bshouty1995learning}. The original formulation for quantum PAC considered the problem of learning a Boolean target function $f$ over length-$n$ bitstrings ($\mathcal{X}=\{0,1\}^{\otimes n}$ and $\mathcal{Y}=\{0,1\}$), where one assumes that there exists an oracle\footnote{In computational learning theory, an oracle is a conceptual tool that serves as a black-box providing access to a desired input-output relationship. A learner queries the oracle by providing an input, and the oracle instantly returns the corresponding correct output. In standard PAC, an oracle produces a pair $(x,f(x))$ with probability $p_\mathcal{D}(x)$.} which outputs superpositions of labeled examples
\begin{equation}
    \ket{\psi_f^{\mathcal{D}}}=\sum_{x} \sqrt{p_\mathcal{D}(x)}\ket{x,f(x)}\,.\label{eq:quantum-example}
\end{equation}
Here, $p_\mathcal{D}(x)$ denotes the probability of sampling the bitstring $x$ with respect to a distribution $\mathcal{D}$ over $\mathcal{X}$. The classical equivalent is an oracle that outputs the mixed state $\sum_{x} p_\mathcal{D}(x)\ket{x,f(x)}\bra{x,f(x)}$, as measuring $\ket{\psi_f^{\mathcal{D}}}$ in the computational basis reduces the quantum oracle to the classical one.

Notably, as proved in~\cite{arunachalam2018optimal}, in distribution-free PAC, if one makes no assumptions regarding $\mathcal{D}$, then for any hypothesis class with VC dimension $d$, quantum and classical sample complexities coincide up to constant factors. That is, quantum examples alone do not improve the worst-case VC-based sample complexity beyond constants. This has since been extended to batch multiclass learning, online Boolean learning, and online multiclass learning~\cite{mohan2025quantum}.
 
The information-theoretic no-separation in the fully general case can fail once one restricts to specific distributions or hypothesis classes \cite{caro2020quantumlearning}. In these regimes, access to quantum examples can help, and larger classical–quantum separations are known (see the box below for a PAC version of the Bernstein–Vazirani problem~\cite{bernstein1993quantum}). A prominent example are disjunctive normal forms (DNFs) under the uniform distribution: DNFs—Boolean formulas that are an OR of ANDs (e.g., $(x_1 \wedge \neg x_3) \vee (x_2 \wedge x_4)$). The latter can be learned efficiently from polynomially many quantum examples under the uniform distribution~\cite{bshouty1995learning}, while no efficient classical algorithm is known for PAC learning DNF in the same setting (the classical learner would need superpolynomial samples and runtime, and is not guaranteed to succeed at all). For $k$-juntas—Boolean functions that depend on at most $k$ of the $n$ input bits (the remaining $n-k$ being irrelevant)—quantum learners under the uniform distribution with superposition oracle access can identify the relevant variables and learn the target with a classical post-processing time that scales more favorably in $k$ and $n$ than the best known classical bounds~\cite{atici2007quantum}. Now, the advantage is primarily in classical computational time complexity rather than in sample complexity. Moreover, when considering $n$-bit Boolean functions that are $k$-sparse in the Fourier domain, and still under a uniform data distribution, there exist regimes ($k\ll n^2$) where the quantum learner uses significantly fewer samples than the best classical procedures~\cite{arunachalam2021two}. Beyond PAC, related property-testing algorithms for $k$-juntas further sharpen separations by exploiting group-theoretic structure (e.g., the quantum Fourier transform over the symmetric group)~\cite{ambainis2016efficient,chen2023testing}. These results showcase the importance of exploiting group harmonic learning techniques, and these insights have even been used to show quantum advantages in learning from noisy data~\cite{cross2015quantum,riste2017demonstration,grilo2019learning} 

\begin{mybox}{Quantum PAC learner for parities via quantum examples (Bernstein--Vazirani as PAC).}

Here we present a simple PAC example which exhibits a quantum advantage for learning a linear parity function under a uniform distribution. Let the input domain be $\mathcal{X}=\{0,1\}^n$, while the labels are  $\mathcal{Y}=\{0,1\}$. The goal is to learn an unknown function from the concept class $\mathcal{F}_{\rm BV}=\{f_s : f_s(x)= s\cdot x \bmod 2,\ s\in\{0,1\}^n\}$. We focus on parities as they are some one of the simplest Boolean functions, are balanced (half of their outputs are 0s, and the other half are 1s) and have a VC-dimension of $n$, thus constituting a nontrivial PAC class. 

To solve this task, one assumes access to copies of the quantum example state 
  \[
    \ket{\psi_{f_s}^{{\rm unif}}} \;=\; \frac{1}{\sqrt{2^n}} \sum_{x\in\{0,1\}^n} \ket{x}\ket{f_s(x)}\,,
  \]
where the first register are the $n$ data qubits, and the second the label qubit. Note also that in the previous equation we have explicitly noted that examples are sampled according to the uniform distribution, and we consider the hypothesis class $\mathcal{HC}_{\rm BV}=\mathcal{F}_{\rm BV}$. That is, learner outputs a parity function. In particular, consider the action of the following unitaries, which outputs the bitstring $s$ from $\ket{\psi_{f_s}^{{\rm unif}}}$ with a sample complexity of $1$,
\begin{align*}
\ket{\psi_{f_s}^{{\rm unif}}} &\xrightarrow{H\text{ on label qubit}} 
 \frac{1}{\sqrt{2^{n+1}}} \sum_{x\in\{0,1\}^n} \ket{x}\ket{0}+(-1)^{f_s(x)}\ket{x}\ket{1}\xrightarrow{H\text{ on data qubits}}  \frac{1}{\sqrt{2}}\left(\ket{0}^{\otimes n}\ket{0}+\ket{s}\ket{1}\right)\,.
\end{align*}
Hence, measuring all qubits in the computational basis and post-selecting on the label qubit being $\ket{1}$ recovers $s$, from which one can output $h=f_s$ (zero generalization error). This yields an $(\varepsilon,\delta)$-PAC learner with quantum sample complexity $1$ and time $O(n)$ gates. Classically, identifying $s$ from i.i.d.\ randomly labeled examples requires $\Omega(n)$ samples in general. Morally, having coherent access to the data and performing interference via a simple Walsh–Hadamard transform allows one to obtain global properties in one go.
\end{mybox}

A complementary route is to focus on computational time complexity rather than sample complexity, by choosing concept classes linked to cryptographic assumptions. It was first noted in~\cite{kearns1994cryptographic} that efficiently PAC learning certain Boolean function classes (under standard assumptions) would allow one to break cryptographic systems (e.g., RSA based on factoring Blum integers). 
Building on this idea, Ref.~\cite{servedio2001quantum} proved the first explicit computational runtime separation. Here, one constructs concept classes that are efficiently learnable quantumly but not classically (unless those assumptions fail). Specifically, one of these constructions  encodes a task that is as hard as factoring for classical learners, while a quantum learner can invoke Shor’s algorithm~\cite{shor1994algorithms} as a subroutine to learn efficiently. Under the assumption that factoring is classically hard, this gives a clear computational separation in an efficient PAC sense. 

Similarly to the previous approach, researchers have identified concept classes where quantum classifiers can solve tasks exponentially faster or more accurately than any classical learner, under reasonable complexity assumptions. 
A notable example is the discrete-logarithm-based classification problem of~\cite{liu2021rigorous}. There, the authors map data into a quantum kernel space such that no classical algorithm can classify better than random guessing (under discrete-log hardness), while proving that there exists a quantum algorithm that attains high accuracy. This separation hinges on the assumption that evaluating the data generating function--i.e., the discrete log--is hard for a classical computer. Crucially, one can also define learning task where the goal is to identifying--rather than evaluating--the functions that generate the data. This is precisely the approach taken in~\cite{gyurik2023exponential}, where the authors prove two learning separations based on the classical hardness of identifying the discrete logarithm and the discrete cube root function.

Going beyond PAC, learning separations have also been proved for active supervised learning tasks within the exact learning with membership queries. Here, the goal is to exactly learn—say with fixed success probability—a Boolean function $f$ over $x\in\{0,1\}^{\otimes n}$ by interactively querying an oracle. Classically, the oracle returns $x_i$ on query $i\in[n]$. In the quantum realm, a query corresponds to $O_x:\ket{i,b}\rightarrow \ket{i,b\oplus x_i}$, with $i$ in binary and $b\in\{0,1\}$. In this framework, the Bernstein–Vazirani problem from the box above collapses to a single quantum query by querying a superposition of all basis vectors, giving an $\Omega(n)$ to $\mathcal{O}(1)$ reduction in query complexity. Despite such examples, general results show that quantum improvements in query complexity are at most polynomial for broad classes~\cite{servedio2004equivalences}, precluding generic query exponential separations. However, if one instead focuses on  time as metric of interest, then Ref.~\cite{servedio2004equivalences} also leverages Shor’s algorithm to exhibit a concept class that is efficiently exactly learnable from membership queries by a quantum learner but not by a classical one. This again highlights the importance of being explicit about which resource—samples/queries or time—is the target of improvement.

Next, we note that in addition to large exponential gaps for special cases, broad but smaller quantum improvements have been established in learning theory by going beyond the strict definition of PAC~\cite{arunachalam2018optimal}. For instance, if  instead of copies of $\ket{\psi_f^{\mathcal{D}}}$, one has access to the circuit $U_f$ that prepares this state, then up to polylogarithmic factors there is a tight square-root improvement over the classical PAC sample complexity~\cite{salmon2024provable}.
This result suggests that, by relaxing the definition of standard PAC and allowing one to peak into the oracle black-box so that the unitary $U_f$ (and its inverse $U_f^\dagger$) can be implemented, one can obtain a provable polynomial improvement in sample complexity for arbitrary concept classes.

To finish, it is worth noting that the framework of quantum computational learning theory has achieved what variational QML has so far failed to do: provide problems where provable separations between classical and quantum learners exist. The caveat is that the problems for which one can prove separations are often artificial, have little-to-no practical application, or ultimately rely on cryptographic assumptions. In a way, one can view this as following from a ``law of conservation of weirdness''~\cite{aaronson2022much}, according to which an exponential separation can only arise if the problem is sufficiently structured--likely to the point of being impractical. In Sec.~\ref{sec:chapQML:qml-linear-algebra} we will study a different approach to QML that was believe to both (i) tackle practical problems and (ii) exhibit exponential separations, and we will discuss whether the law of conservation of weirdness still holds there.

\subsection{Learning quantum concepts}\label{sec:chapQML:PAC-quantum}

In the previous section we discussed provable ways in which quantum learner can more efficiently solve problems on classical data. There, training examples are ordinary bitstrings or real-valued vectors, meaning that the quantum learner does not necessarily gain an information-theoretic head start--it ultimately sees the same labeled bits a classical learner does--but it can sometimes process them more efficiently.  With quantum data the story is different. Now the examples may be quantum states or general quantum processes. Comparing classical and quantum learners becomes subtler because objects that live natively on a quantum device (states, channels) rarely admit compact classical representations without losing structure. 

As we will see below, the tools from learning theory have significantly reshaped how information is extracted from quantum experiments. Indeed, the task of learning from objects living in quantum computers is quite subtle as access is governed by the Born rule, one needs to deal with the fact that many observables are incompatible (non-commuting), measurements are destructive, copies are finite (no-cloning), as well as with shot noise uncertainties and additional hardware error sources. Hence, when talking about learning quantum concepts one tracks laboratory-relevant resources: how many uses of the state or process (fresh copies, channel calls), whether collective (entangled) measurements are allowed or one must measure copy-by-copy, whether quantum memory can be maintained across rounds, and the circuit depth/width of the learning routine. Framing these constraints explicitly as learning problems has led to protocols with provable guarantees and practical wins, making learning from quantum data one of QML’s clearest successes in day-to-day laboratory work.

Perhaps the simplest and most informative example of how quantum data can lead to new learning tasks is the problem of learning a quantum state (see  Refs.~\cite{anshu2024survey,gebhart2022learning} for a survey on this topic). Here, the goal is to produce a model capable of predicting outcomes for either all--or some restricted set--of measurements over the state. As always, care must be taken in how we define what ``learning'' means as different definitions will lead to different computational complexions. Here we will discuss two forms of learning: state tomography, and PAC learning a quantum state.

Quantum tomograph refers to the problem of obtaining--from measurements--a classical description of an unknown quantum state $\rho$. In particular, success is determined if the classical description, denoted as $\widehat{\rho}$, is $\varepsilon$ close $\rho$ in  trace distance metric. For convenience, we recall that given two states $\rho$ and $\sigma$, their trace distance $T(\rho,\sigma)=\tfrac{1}{2}\|\rho-\sigma\|_1$ serves as an upper bound for the difference between the measurement outcome of any positive operator $E$~\cite{nielsen2000quantum}. That is
\[
\max_{0\leq E\leq I}\,{\rm Tr}\big[(\rho-\sigma)E\big]=T(\rho,\sigma)\,.
\]
Hence, one can guarantee that no measurement outcomes can differ by more than $\varepsilon$ if $T(\rho,\widehat\rho)\leq \varepsilon$. 

Given that a general $n$-qubit state is specified by $4^n-1$ real parameters, full tomography has information-theoretic sample requirements that grow at least like $\Omega(4^n/\varepsilon^2)$ to reach accuracy $\varepsilon$~\cite{haah2017sample,o2016efficient}. The factor $4^n$ can be improved to $2^n r$  when learning a state that is promised to have rank $r$~\cite{haah2017sample,o2016efficient,gross2010quantum,cramer2010efficient}.  Thus, even with structure and clever reconstruction, optimal tomographic routines become impractical very quickly as the number of qubits grows. 

While this exponentially large sample complexity allows one to fully learn arbitrary quantum states and predict any measurement outcome, the situation changes if we instead restrict to learning scenarios where one has a promise that the unknown state belongs to a restricted subset of states. In the box below, we explicitly work out a tomography pipeline for learning a generic tensor product state. In addition,  results in the literature have been obtained for learning other families of states. These include  learning $n$-qubit stabilizer state, i.e., states prepared by initializing a quantum computer to the all zero state $\ket{0}^{\otimes n}$ and evolving it with a Clifford unitary~\cite{tolar2018clifford,mastel2023clifford}~\footnote{Here we briefly recall that Clifford circuits are a special subset of unitaries composed only of Hadamard, $S$ phase gates,  and CNOTs. They play an important role in quantum computing as--for certain input states--quantum
circuits composed of Clifford gates can be efficiently simulated classically via the Gottesman-Knill theorem~\cite{gottesman1998heisenbergrepresentation,aaronson2004improved,nest2008classical}.}. With single copy access requires a number of samples scaling as $\mathcal{O}(n^2)$~\cite{aaronson2004improved}. If instead one can coherently access two copies of the state and make entangled measurements, the sample complexity reduces to $\mathcal{O}(n)$~\cite{montanaro2017learning} and one can obtain a time complexity of $\mathcal{O}(n^2)$. Another example is that of learning Gaussian, or free fermions, states. These states are obtained by preparing the all zero state $\ket{0}^{\otimes n}$ and evolving it with a free-fermionic unitary\footnote{We also recall that free fermionic circuits are a special subset of unitaries taking the form $U=e^{\sum_{i<j}h_{ij}\gamma_i \gamma_j}$ for $h_{ij}\in\mathbb{R}$, and where we defined the set of $2n$ Majoranas $\{\gamma_i\}_{i=1}^{2n}$ as $    \gamma_1=XI\dots I$, $    \gamma_2=YI\dots I$, $    \gamma_3=ZX\dots I$, $    \gamma_4=ZY\dots I$, $\cdots$, $    \gamma_{2n-1}=ZZ\dots X$, $    \gamma_{2n}=ZZ\dots Y$. Just like Clifford circuits, for certain input states free fermionic circuits can be efficiently simulated~\cite{jozsa2008matchgates,valiant2001quantum, knill2001fermionic,terhal2002classical,goh2023lie}. }. As shown in~\cite{gluza2018fidelity,aaronson2021efficient,o2022fermionic,mele2024efficient}, one can learn free fermionic states in with a sample and time complexity that scales polynomially with the number of fermions and the number of modes in the system. Finally, matrix product state~\cite{cramer2010efficient,lanyon2017efficient}, finitely correlated states~\cite{fanizza2023learning}, high-temperature Gibbs states~\cite{rouze2024learning}, states prepared with shallow circuits~\cite{landau2025learning,kim2024learning,huang2024learning} are also known to be efficiently learnable.

\begin{mybox}{Learning an $n$-qubit product state by state tomography}

In this box we consider a simple learning task where the goal is to learn an unknown $n$-qubit product state
\[
\rho = \rho_1\otimes \rho_2\otimes \cdots \otimes \rho_n\,,
\]
so that one can predict any observable over $\rho$. In what follows, we denote  $\mu_{j,a}=\mathrm{Tr}(\rho_j\,\sigma_a)$ for $j\in[n]$ and $a\in\{X,Y,Z\}$. 

The tomographic data is experimentally collected by taking $3N$ copies of $\rho$, and measuring all qubits in parallel in the $X$, $Y$ and $Z$ basis. For each qubit, one estimates its Bloch vector components $\widehat\mu_{j,a}$, which describes the learned state $\widehat\rho=\bigotimes_j \widehat\rho_j$ with $\widehat\rho_j=\frac{1}{2}\left(I+ \widehat\mu_{j,X} X+ \widehat\mu_{j,Y} Y+ \widehat\mu_{j,Z} Z\right)$. 

The trace distance between $\rho$ and $\widehat\rho$ can be bounded as follows:
\[
T\left(\rho,\widehat\rho\,\right)
\le
\sqrt{1-F(\rho,\widehat\rho)}=\sqrt{1-\prod_{j=1}^nF(\rho_j,\widehat\rho_j)}\leq\sqrt{1-\prod_{j=1}^n\left(1-T(\rho_j,\widehat\rho_j)\right)^2}\leq\sqrt{\sum_{j=1}^nT(\rho_j,\widehat\rho_j)(2-T(\rho_j,\widehat\rho_j))}\leq\sqrt{2\sum_{j=1}^nT(\rho_j,\widehat\rho_j)}\,,
\]
where we have defined the quantum fidelity $F(\rho,\sigma)={\rm Tr}\left[\sqrt{\sqrt{\rho}\sigma\sqrt{\rho}}\right]^2$, and where we have used the inequalities $1-\sqrt{F(\rho,\sigma)}\leq T(\rho,\sigma)\leq\sqrt{1-F(\rho,\sigma)}$,
the fact that fidelity is multiplicative
with respect to tensor-product states  and the inequality $\prod_j (1-d_j)\geq 1-\sum_j d_j$ for $d_j\in[0,1]$ and with $d_j=2T(\rho_j,\widehat\rho_j)-T(\rho_j,\widehat\rho_j)^2$. Next, for single qubits, the trace distance equals half the Euclidean distance of the Bloch vectors
\[
T(\rho_j,\widehat\rho_j)=\tfrac{1}{2}\big\|\left(\mu_{j,X},\mu_{j,Y},\mu_{j,Z}\right)-\left(\widehat\mu_{j,X},\widehat\mu_{j,Y},\widehat\mu_{j,Z}\right)\big\|_2
\;\le\;\tfrac{\sqrt{3}}{2}\,\max_{a\in\{X,Y,Z\}}\big|\,\mu_{j,a}-\widehat\mu_{j,a}\big|.
\]
From here, we can study how many samples are needed to guarantee that 
$\big|\widehat\mu_{j,a}-\mu_{j,a}\big|\le\widetilde{\varepsilon}
$ via Hoeffding's inequality
\[
\Pr\Big[\,|\,\widehat\mu_{j,a}-\mu_{j,a}|\ge \widetilde{\varepsilon}\,\Big]\;\le\;2\,e^{-2N\widetilde{\varepsilon}^2}\,.
\]
Applying a union bound over all $3n$ coordinates $(j,a)$ gives
\[
\Pr\Big[\,\max_{j\in[n],\,a\in\{X,Y,Z\}}|\,\widehat\mu_{j,a}-\mu_{j,a}|\ge \widetilde{\varepsilon}\,\Big]
\;\le\; 6n\,e^{-2N\widetilde{\varepsilon}^2}\,.
\]
Hence, to ensure, with probability at least $1-\delta$, that $|\,\widehat\mu_{j,a}-\mu_{j,a}|\le \widetilde{\varepsilon}$ for all $(j,a)$ simultaneously it requires $N=\frac{1}{2\widetilde{\varepsilon}^2}\log\left(\frac{6n}{\delta}\right)$, which implies $\widetilde{\varepsilon}=\sqrt{\frac{1}{2N}\log\left(\frac{6n}{\delta}\right)}$. 

It follows that $T\left(\rho,\widehat\rho\,\right)\leq \varepsilon$ if  $\sqrt{\frac{3n}{2N}\log\left(\frac{6n}{\delta}\right)}\leq \varepsilon$, from where one can find
\[N\geq \frac{3n}{2\varepsilon^2}\log\left(\frac{6n}{\delta}\right)\,.\]

Since each round uses one fresh copy of the $n$-qubit state $\rho$, with the choice of $N$ above our algorithm requires a number of copies scaling as $N\in\mathcal{O}\Big(\tfrac{n}{\varepsilon^{2}}\log\left(\tfrac{n}{\delta}\right)\Big)$. Note that this algorithm allows one to learn a general product state because such states are fully specified by only the $3n$ single-qubit Pauli means. For the family of tensor product states, no global tomography is needed, and the learner directly outputs a valid product-state hypothesis $\widehat\rho$.
\end{mybox}

The problem of learning a quantum state can also be cast in the PAC framework. Whereas tomography aims to predict every measurement outcome, PAC learning seeks to predict the outcomes of ``most'' measurements. As such, one considers datasets drawn from a fixed, restricted family of measurements and aims to predict selected properties of the state. Now, a general mixed state $\rho$ can be regarded as a map from the input domain of positive operator-valued measures (POVMs) elements $\mathcal{X}\subseteq\{E\succeq 0:\,E\le\mathbbm{1}\}$  to the label domain $\mathbb{R}$\footnote{Note that in the literature it is standard to consider  $\mathcal{X}$ to instead be the set of two-outcome POVMs. That is, $E$ is hermitian with eigenvalues in $[0,1]$. This set is informationally complete, in the sense that known ${\rm Tr}[\rho E]$ for all such $E$ is equivalent to fully knowing $\rho/$ }. That is, $f_\rho:\mathcal{X}\to[0,1]$, where $f_\rho(E)={\rm Tr}[\rho E]$. Training sets then take the form $\mathcal{S}=\{(E_i,
{\rm Tr}[\rho E_i])\}_{i=1}^N$, where the POVMs were sampled according to some probability distribution $\mathcal{D}$,  and the task is to find a hypothesis $h\in\mathcal{HC}$ that generalizes to unseen measurements~\cite{aaronson2007learnability}. As we can see, in the language of PAC, $\rho$ becomes the target unknown concept that one wishes to learn. 

The first notable result in PAC learning of quantum states was obtained in~\cite{aaronson2007learnability}. Therein, it was shown that unlike full state tomography which requires an exponentially large number of samples to learn a generic state, PAC learning requires a number of samples scaling only as $\mathcal{O}(n)$. While this constitutes an exponential reduction in the sample complexity, the catch is that the time complexity of the problem one needs to solve to learn the state is still exponential in $n$. As before, by focusing on special families of states, one can study if whether they are both sample- and time-efficiently PAC-learnable. In the box below we showcase how the number of samples changes when learning a tensor product state. Then, Ref.~\cite{rocchetto2017stabiliser} showed that if $\mathcal{X}$ constitutes Pauli observables, then stabilizer
states are PAC-learnable in polynomial time. Importantly, we note that while tensor product states and stabilizers show  that PAC learning is somewhat easier than tomography, this is not a generally applicable result. As a counter example to such a claim, learning Gaussian states has been shown to be NP-hard~\cite{bittel2025pac}, thus evidencing that how hard or easy to it is to solve a learning task over a quantum state, truly depends on the definition of that we mean by ``learn'' and that classical simulability might not always entail efficient learnability~\cite{yoganathan2019condition}.

\begin{mybox}{Learning an $n$-qubit product state in a PAC-style setup}

In this box we consider a simple PAC-style learning task where the goal is to learn an unknown product state
\[
\rho \;=\; \rho_{1}\otimes \rho_{2}\otimes \cdots \otimes \rho_{n}\,,
\]
so that we can predict any observable over $\rho$. Here, the input domain $\mathcal{X}$ consists of a pair $(j,a)$ where $j\in[n]$ and $a\in\{X,Y,Z\}$, and we define the exact local expectation values as $\mu_{j,a}=\mathrm{Tr}(\rho_j\,\sigma_a)$. In an experimental setting, we can collect data by preparing a fresh copy of the target $n$-qubit state $\rho$, picking an axis $a$ and measuring all qubits in basis $a$, producing labels $y_{j,a}^{(i)}\in\{-1,+1\}$ for every $j$. This scheme thus produces $n$ labeled outcomes per round (one for each $(j,a)$ present in that round). 
The hypothesis class consists of product states $\widehat\rho=\bigotimes_j \widehat\rho_j$ with $\widehat\rho_j=\frac{1}{2}\left(I+ \widehat\mu_{j,X} X+ \widehat\mu_{j,Y} Y+ \widehat\mu_{j,Z} Z\right)$ determined by their Bloch vectors $\left(\widehat\mu_{j,X},\widehat\mu_{j,Y},\widehat\mu_{j,Z}\right)$.

The learning procedure is as follows. Given a dataset obtained from $3N$ rounds of measurements—where at each round we cyclically choose $a\in\{X,Y,Z\}$—we set $\widehat\mu_{j,a}$ to the empirical average of the outcomes collected for that pair $(j,a)$. That is, $\widehat\mu_{j,a}=\frac{1}{N}\sum_{i=1}^N y_{j,a}^{(i)}$. By Hoeffding's inequality, and applying a union bound over all $3n$ coordinates $(j,a)$ we obtain that it suffices to choose
\[
N\ge\frac{1}{2\varepsilon^2}\log\left(\frac{6n}{\delta}\right)\,,
\]
to ensures that, with probability at least $1-\delta$, $|\widehat\mu_{j,a}-\mu_{j,a}|\le\varepsilon$ for all $(j,a)$ simultaneously. In particular,
$\mathcal{L}(\widehat\rho)=\mathbb{E}_{(j,a)}[(\widehat\mu_{j,a}-\mu_{j,a})^2]\le \varepsilon^2$. 
As we can see, the number of samples needed in this PAC-style learning setup is an order of $n$  smaller than those needed to perform tomography. 
\end{mybox}

Beyond tomography and PAC learning, we can frame the problem of learning as a property testing task. Here, the goal is to use measurements to  decide  whether a state belongs to a given class, or is significantly distant from any state belonging to that class. We refer the reader to~\cite{montanaro2013survey} for a survey on this topic. More recently, the framework of classical shadows~\cite{huang2020predicting, aaronson2019shadow,elben2022randomized,paini2019approximate} has attracted considerable attention as a practical and experimentally feasible tool to learn a quantum state. Here, rather than reconstruct the entire density matrix, or estimating ``most'' observables, we seek to estimate specific properties we care about. In a nutshell, a classical shadow protocol consists of sending the state $\rho$ we wish to learn through a random unitary $U$ sampled from a given unitary group (e.g., random local single-qubit rotation, random free-fermionic unitary, etc) and measuring in the computational basis, thus obtaining a bitstring $\ket{z}$ with $z\in\{0,1\}^{\otimes n}$. The tuple $\{U,\ket{z}\}$ is called a ``snapshot'', and can be used to reconstruct the target state as $\rho=\mathbb{E}[\mathcal{M}^{-1}(U^\dagger \ket{z}\bra{z}U)]$ where $\mathcal{M}$ is the channel mapping from $\rho$ to its 
snapshot and the expectation value is taken over the random unitaries and the measurement outcomes. The specific form of $\mathcal{M}$, as well as the sets of observables which can be efficiently estimated depend on the choice of group from which $U$ is sampled~\cite{wan2022matchgate,buadescu2021improved,zhao2021fermionic,hu2021classical,low2022classical,van2022hardware,koh2022classical,chen2021robust,hearth2024efficient,sauvage2024classical,west2025real,bertoni2022shallow}. In the box below, we showcase a specific example. If the learning task needs many expectation values, shadows give an information-theoretically near-optimal way to get them without ever building a full tomographic model~\cite{huang2021information}.

\begin{mybox}{Classical shadows with local random unitaries.}

Given an unknown $n$-qubit state $\rho$, our goal is to estimate the expectation value of $L$ Pauli observables without full tomography by using classical shadows. Each snapshot is produced by (i) sampling a local unitary $U=\bigotimes_{j=1}^n U_j$ from the single-qubit Clifford group (a unitary 3-design, so we can use Cliffords instead of Haar random unitaries), (ii) applying $U$ to $\rho$, and (iii) measuring in the computational ($Z$) basis to obtain a bitstring $z\in\{0,1\}^n$. Each time we run the experiment, we obtain a snapshot  $(U,\ket{z})$.

Let $\mathcal{M}$ denote the measurement channel that maps $\rho$ to its snapshot, one can explicitly show that~\cite{huang2020predicting} $
\mathcal{M}(\rho)\;=\;\mathbb{E}_{U,z}\!\left[\,U^\dagger \ket{z}\!\bra{z}\,U\,\right]
\;=\;\bigotimes_{j=1}^n \mathcal{M}_1(\rho)$ with $\mathcal{M}_1(X)=\frac{X+I}{3}$. Hence, its inverse acts locally on each qubit, yielding the shadow estimator
\[
\widehat{\rho}\;=\;\mathcal{M}^{-1}\!\big(U^\dagger \ket{z}\!\bra{z}\,U\big)
\;=\;\bigotimes_{j=1}^n \Big(3\,U_j \ket{z_j}\!\bra{z_j}\,U_j^\dagger - I\Big)\,, \quad \text{where}.
\]
By construction, $\mathbb{E}_{U,z}[\widehat{\rho}]=\rho$, meaning that $\widehat{\rho}$  is an unbiased estimator of the target state. 

Given $N$ i.i.d.\ snapshots $\{(U^{(i)},\ket{z^{(i)}})\}_{i=1}^N$, we estimate an observable $O$ by the sample mean
\[
\widehat{o}\;=\;\frac{1}{N}\sum_{i1}^N \mathrm{Tr}\!\left[\,O\,\widehat{\rho}^{(i)}\right],\qquad 
\widehat{\rho}^{(i)}=\mathcal{M}^{-1}\!\big((U^{(i)})^\dagger \ket{z^{(i)}}\!\bra{z^{(i)}}\,U^{(i)}\big).
\]
For a family $\{O_\ell\}_{\ell=1}^L$ of Pauli observables that are at most $k$-local, the variance/shadow-norm analysis gives the sample bound~\cite{huang2020predicting}
\[
N \; \geq\!\frac{3^{\,k}}{\varepsilon^2}\,\log\left(\frac{2L}{\delta}\right)\,,
\]
to ensure, with probability at least $1-\delta$, that all estimates satisfy $|\widehat{o}_\ell-\mathrm{Tr}(O_\ell\rho)|\le \varepsilon$ simultaneously. As one can see, classical shadows with local random unitaries enable the efficient estimation of expectation values for low-bodyness Paulis. Intuitively, the $3^{k}$ factor reflects that local Cliffords depolarize each qubit by $1/3$, and only the $k$ active qubits of a $k$-local Pauli contribute to the variance.
\end{mybox}

The theory of quantum learning can also be applied to study the complexity of learning unknown quantum processes~\cite{huang2022foundations,jerbi2023power}. Now we can define learning, for instance, as being able to implement the process at will. This includes problems such as learning generic channels~\cite{d2001quantum,wadhwa2023learning,wadhwa2024agnostic} or special cases of unitary~\cite{huang2024learning,bisio2010optimal,sedlak2019optimal,garcia2023deep,angrisani2023learning,leone2024learning,austin2025efficiently,zhao2024learning,vasconcelos2024learning}, Hamiltonian Learning~\cite{wiebe2014quantum,wiebe2014hamiltonian,gu2022practical,caro2022learning,huang2022learningmanybodyhamiltonians,yu2023robust,hu2025ansatz,dutkiewicz2023advantage,bakshi2024learning,abbas2025nearly}, process tomography via shadows~\cite{levy2024classical,kunjummen2023shadow,fanizza2024learning,castaneda2025hamiltonian}, learning Pauli or unitary channels~\cite{flammia2021pauli,chen2022quantum,raza2024online,grewal2025query}; with some of those tasks being explicitly formulated as tomographic problems~\cite{wadhwa2024agnostic, mohseni2008quantum} or within the PAC framework~\cite{chung2021sample}\footnote{More than simply a theoretical question,  the ability to learn quantum processes has extensive practical applications. For instance, these tools can be used to mitigate noise in quantum computers. Indeed, knowledge of a noisy channel can enable the implementation of its inverse and reduce the amount of error in quantum computers~\cite{temme2017error,berg2022probabilistic,guimaraes2023noise}.}.

An interesting realization is that across most, if not all, tasks of learning over quantum concepts, allowing collective access--either entangled inputs to a channel, joint measurements across multiple copies, keeping quantum memory across rounds, or being able to prepare a concept (state or unitary) and its complex conjugate--often qualitatively changes and improves the sample complexities~\cite{huang2021quantum,king2024triply,aharonov2021quantum,chen2021exponential,chen2021hierarchy,king2024exponential,sharma2020reformulation,seif2024entanglement,oh2024entanglement}. This again showcases how different access models can change the learning complexity, and reveal a very rich landscape of experimental setups than one can use in practice.

Next, we note that some of the tools originally developed to learn quantum concepts have already been used to learn classical concepts. This includes, among others, learning classical instances with quantum labels~\cite{caro2021binary}, and tasks where classical shadows can be used to obtain exponential separations between classical and quantum learners ~\cite{jerbi2023shadows}. Indeed, this cross-pollination of ideas is already allowing researchers to improve and enhance our understanding of quantum computational learning theory.

In short, learning from quantum data is where QML has had its clearest practical impact to date. The interplay between information-theoretic limits (copy complexity, measurement incompatibility, disturbance) and algorithmic design (adaptive measurement, shadow-based estimators, compressed sensing, Hamiltonian/process learning) has translated into tools used daily in laboratories. This point of view—treating experimental design and inference as a learning problem--both clarifies what is and is not possible and continues to deliver sample- and time-efficient procedures for real devices which ultimately translate into valuable time and resource saving for experimentalists.

\section{Quantum machine learning from linear-algebraic primitives}
\label{sec:chapQML:qml-linear-algebra}

In this section we go back to learning tasks over classical data. As discussed above, the variational QML framework can be used to tackle real-world problems on classical data, but offers few end-to-end guarantees. On the other hand, quantum computational learning theory delivers guarantees for stylized tasks under strong assumptions, but these do not necessarily resemble problems one encounters on a day-to-day basis. In this section, we turn to a slightly different approach to QML: using quantum algorithmic primitives to enhance ML tasks. The motivation is that many well-established techniques--such as linear regression, PCA, and singular value decomposition--are built on linear-algebraic transformations and serve as essential computational workhorses across a wide range of applications in ML and data analysis. Hence, there was a big push in the mid-2010s to ask whether implementing these transformations on quantum hardware--acting on amplitude-encoded data or via block-encodings of matrices--could both tackle practical tasks and yield exponential speedups. As we will see, initial separations were claimed, but subsequent work dequantized many of these algorithms by providing classical methods with comparable access models and resource requirements.

Ultimately, linear-algebraic QML is a cautionary tale about data access. The headline speedups reported in the mid-2010s hinged on strong oracles for loading and manipulating classical data (e.g., quantum random access memory (QRAM)/state preparation, $\ell_2$-sampling, block-encodings). Once analogous access is granted to classical algorithms, similar performance can often be recovered, leading to dequantization. To make the quantum data-access model concrete, we present in the box below a quantum binary search tree (BST) that supports coherent, logarithmic-time amplitude preparation, in the spirit of QRAM~\cite{giovannetti2008quantum, jaques2023qram}. In a nutshell, QRAM stores the classical data in a structure that allows to read data at once via quantum superposition. Given a classical memory $T$, it implements a function $\mathcal{R}$ such that
\[ \mathcal{R}(T): \ket{i}\ket{0} \to \ket{i}\ket{T_i}\;, \]
where $\ket{i}$ defines the memory address, and $\ket{T_i}$ the data stored at that address~\cite{jaques2023qram}\footnote{Here, we restrict our attention to classical memory, but similar quantum access is possible with quantum memory $\ket{T_0, \dots T_{N-1}}$, represented as $\mathcal{R}: \ket{i}\!\ket{0}\!\ket{T_0,\dots, T_{N-1}}  \to \ket{i}\!\ket{T_i}\!\ket{T_0,\dots, T_{N-1}}$~\cite{jaques2023qram}}. This abstraction enables to access to a superposition of multiple memory address states, and load the classical data into the quantum states simultaneously.

\begin{figure}
    \centering
    \includegraphics[width=1\linewidth]{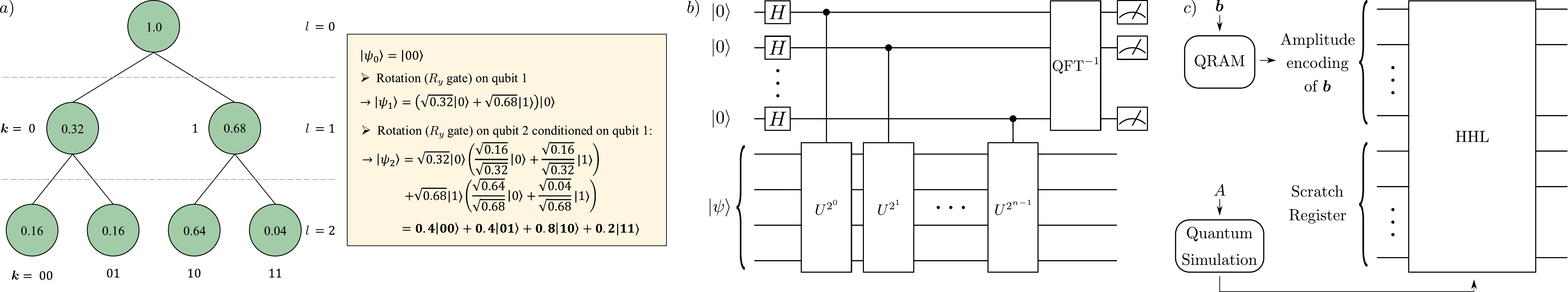}
    \caption{a) Example of a binary tree $\mathsf{B}_{i}$ to store a classical state $\boldsymbol{x}$ and to prepare the data structure for a two-qubit state. The algorithm on the right illustrates how the classical vector stored in the binary tree (left) is encoded into a quantum state.  b) Circuit for the quantum phase estimation algorithm applied to a unitary operator $U$ via $n$ ancilla qubits. Here, QFT denotes the quantum Fourier transform~\cite{nielsen2000quantum} and QFT$^{-1}$ its inverse. c) Circuit for using the Harrow-Hassidim-Lloyd (HHL) algorithm to tackle the quantum linear system problem to solve $A \boldsymbol{x} = \boldsymbol{b}$. The first register is used to load the input vector $\boldsymbol{b}$ and the resulting solution state, while the second scratch register provides additional qubits required for QPE algorithm.  Both QRAM and the ability to perform quantum simulation on $A$ to implement $e^{iA}$ are basic ingredients of this technique.}
    \label{fig:chapQML:bst}
\end{figure}

\begin{mybox}{Quantum Data structure: Binary Tree}

An efficient data access model is a prerequisite for achieving quantum speedups in linear-algebraic algorithms.
A widely used conceptual architecture is the binary tree  data structure~\cite{kerenidis2016quantum}, which allows polylogarithmic-time access to the entries of a classical matrix $A \in \mathbb{R}^{m\times n}$ and supports efficient amplitude encoding of its rows and columns as quantum states. 
The classical data is stored in a collection of $m$ binary trees $\{\mathsf{B}_i\}_{i=1}^m$ where each binary tree $\mathsf{B}_i$ encodes the entries of the normalized row vector $A_i \in \mathbb{R}^n$. Each leaf node of $\mathsf{B}_i$ stores the squared amplitude $A_{ij}^2$ along with the sign of the corresponding element, while each internal node stores the sum of its two child nodes. During quantum state preparation, the circuit performs a sequence of controlled rotations guided by the tree structure. At each level, the rotation angles are determined by the relative weights of the child nodes, distributing amplitudes recursively from the root to the leaves.

To describe this process more concretely, let $B_{i,\boldsymbol{k}}$ denote a node in the binary tree $\mathsf{B}_i$, identified by the binary string $k \in \{0,1\}^l$, where $l = 1, \dots, \lceil \log(n) \rceil$ denotes the depth of the leaf nodes and the root corresponds to height~$0$. 
Child nodes are indexed such that $0$ corresponds to the left branch and $1$ to the right branch. 
Given a binary tree $\mathsf{B}_i$, QRAM implements the unitary mapping as 
\[\mathcal{R}(\mathsf{B}_i): \ket{i}\!\ket{0} \to \ket{i}\ket{\mathsf{B}_i}\;,\] 
with $\ket{\mathsf{B}_i}$ encoding the (normalized) row elements of the matrix. 
More explicitly, state-preparation procedure evolves recursively following the tree from root to the leaf as
\begin{align}
\ket{\psi_0} = \ket{0}^{\otimes L} & \to \ket{\psi_1} =  \left(\sqrt{B_{i, 0}}\ket{0} + \sqrt{B_{i, 1}} \ket{1}\right)\ket{0}^{\otimes (L - 1)} \nonumber \\ 
& \to \ket{\psi_2} =\left(\sqrt{{B_{i, 0}}} \left(\frac{\sqrt{B_{i, 00}}}{\sqrt{B_{i, 0}}} \ket{00} + \frac{\sqrt{B_{i, 01}}}{\sqrt{B_{i, 0}}} \ket{01}\right) + \sqrt{{B_{i, 1}}} \left(\frac{\sqrt{B_{i, 10}}}{\sqrt{B_{i, 1}}} \ket{10} + \frac{\sqrt{B_{i, 11}}}{\sqrt{B_{i, 1}}} \ket{11}\right) \right)\ket{0}^{\otimes  (L - 2)}  \nonumber \\  
\cdots & \to \ket{\psi_{L}} = \sum_{\boldsymbol{k} \in \{0, 1\}^{L} } \sqrt{B_{i, k}}\ket{k} := \ket{\mathsf{B}_i}\;, 
\end{align} 
where $L= \lceil \log(n) \rceil$ is the total height of the tree and $\{\ket{k}\}$ correspond to the computational basis states. This recursive construction successively distributes amplitudes according to the cumulative values stored in the tree nodes, ensuring that the final quantum state encodes the desired normalized row. Figure~\ref{fig:chapQML:bst}(a) illustrates an example of a binary tree with four leaves, where each level corresponds to a layer of controlled rotations contributing to the final amplitude-encoded state.

The total runtime required to prepare the binary tree structure scales with the depth of the tree, which is at most $\left\lceil \log(n) \right\rceil$, leading to the complexity of $\mathcal{O}(\log^2(mn))$ with memory scaling as $\mathcal{O}(\log^2(mn))$ to build and access the data structure (assuming that the rotations require constant cost). Thus, such a classical data structure with quantum access ensures the linear-algebraic primitives to achieve exponential speed up. However, it is important to note that the binary tree model (and, more generally, most qRAM-based approaches) remains primarily a theoretical model, as constructing and maintaining such architectures with limited error rates poses significant challenges for current hardware~\cite{jaques2023qram}. 

\end{mybox}

A more detailed discussion on the quantum data access model and its connection to dequantization is presented in Sec.~\ref{subsec:dequantization-bridge}.

\subsection{Quantum phase estimation and quantum linear-system solvers}
\label{subsec:hhl}

As we will see below, one of the defining properties of the methods in this section is that they are all based on basic algorithmic textbook techniques, mixing and matching them in different ways to process data. In this section we present two such basic tools: the quantum phase estimation (QPE) algorithm and algorithms for solving linear systems of equations on quantum computers.

The QPE algorithm~\cite{nielsen2000quantum, kitaev1995quantum}, stands out as one of the most fundamental subroutines used in many of the quantum primitives. The goal of QPE is to estimate the eigenvalue $\lambda$ corresponding to a known eigenstate $\ket{v}$ of a unitary operator $U$. Because any unitary operator has eigenvalues of unit modulus, i.e., $\| \lambda\| = 1 $, each eigenvalue can be expressed in exponential form of a real-valued phase $\theta$ as $\lambda = e^{i2\pi \theta}$ with $0 \le \theta < 1$. Hence, estimating the eigenvalue of $U$ is equivalent to determining the eigenphase $\theta$ associated with this exponential representation, giving rise to the name phase estimation. In practice, the estimated phase $\tilde{\theta}$ represents an $n$-bit approximation of the real phase $\theta$ with an error bounded by $|\theta_j  - \tilde{\theta}_j| \le \epsilon$ where $\theta_j$ and $\tilde{\theta}_j$ denotes $j$-th bit of $\theta$ and $\tilde{\theta}$, respectively. The QPE algorithm, shown in Fig.~\ref{fig:chapQML:bst}(b), can be performed with a success probability of $1 - 1/\textrm{poly}(n)$ (for $n$ qubits) and the total runtime scaling as $\mathcal{O}(T_U \log(n)/\epsilon)$, where $T_U$ corresponds to the time required to implement the unitary operator $U$~\cite{kitaev1995quantum}. Therefore, under the assumption that $U$ can be efficiently implemented (in polynomial time), QPE can efficiently estimate eigenphases with the runtime that grows logarithmically with $n$. 

Moving on, let us recall that solving systems of linear equations plays a central role in many classical algorithms across scientific computing, optimization and ML.  
Efficient methods, such as Gaussian elimination, Cholesky decomposition, conjugate gradients have been proposed to speed up many data-driven and ML tasks, especially with the rapidly increasing datasets that define the system size~\cite{golub2012matrix, saad2003iterative}. 
These algorithms form the backbone of numerous data analysis and learning frameworks, including linear regression, SVMs, PCA~\cite{hofmann2008kernel, bishop2006pattern}. 
Quantum computing offers an alternative framework for solving such problems via the so-called quantum linear systems problem (QLSP)~\cite{harrow2009quantum, dervovic2018quantum}. The QLSP is defined as follows. Given a Hermitian matrix\footnote{Although this section only constrains the explanation for the Hermitian matrix, the original HHL algorithm also encompasses the procedure solving non-Hermitian matrices (see Ref.~\cite{harrow2009quantum}). } $A \in \mathbb{R}^{d \times d}$, and a normalized vector $\boldsymbol{b} \in \mathbb{R}^{d}$, one seeks to determine the vector $\boldsymbol{x}$ such that $A \boldsymbol{x} = \boldsymbol{b}$. In the quantum setting, this can be recasted into the quantum states such that $A \ket{\boldsymbol{x}} = \ket{\boldsymbol{b}}$ with $\ket{\boldsymbol{b}} = \frac{1}{d}\sum_{i} b_i \ket{i}$. Since $A$ is Hermitian, it admits a spectral decomposition with eigenstates $\{\ket{u_i}\}$ and corresponding eigenvalues $\lambda_i$. From here, we can expand  $\ket{\boldsymbol{b}}$ and the expected solution $\ket{\boldsymbol{x}}$ in terms of $\{\ket{u_i}\}$
 \begin{equation}
    A = \sum_j \lambda_j \ket{u_j}\bra{u_j},~~~\ket{\boldsymbol{b}} = \sum_{j} \beta_j \ket{u_j}\;, ~~~\ket{\boldsymbol{x}} = \sum_{j}\lambda_j^{-1} \beta_j \ket{u_j}\;, \quad \text{with}\quad \beta_j  = \langle u_j | \boldsymbol{b} \rangle\,. 
    \label{eq:HHL_eq}
\end{equation}

The Harrow-Hassidim-Lloyd (HHL) algorithm~\cite{harrow2009quantum, zaman2023step} has been proposed to provide a quantum procedure to efficiently solve this QLSP with a potential exponential speedup over the classical algorithms (see Fig.~\ref{fig:chapQML:bst}(c) for a circuit). This speed-up stems from the use of QPE to determine the eigenphases of the unitary $U = e^{i A}$, implemented through Hamiltonian simulation techniques~\cite{berry2015hamiltonian}. By definition, the unitary $U$ shares the same eigenstates $\{\ket{u_j}\}_j$ as $A$, with corresponding eigenvalues $e^{i\lambda_j}$. Therefore, estimating the eigenphases of $U$ is equivalent to estimating the eigenvalues of $A$, and the overall accuracy of the HHL algorithm depends directly on the precision of Hamiltonian simulation. As such, to guarantee both efficiency, HHL algorithm makes three main assumptions~\cite{aaronson2015read}: (i) there exists a unitary that can prepare $\ket{\boldsymbol{b}}$, (efficient data access model assumption) (ii) $A$ is $s$-sparse\footnote{This sparsity assumption ensures that the Hermitian matrix $A$ is efficiently simulated. In this regime, the total time complexity for implementing $U(t) = e^{iAt}$ with an arbitrary $t$ scales linearly with respect to $s$. }, meaning that each row and column in the matrix has at most $s$ non-zero entries with $s \ll d$ (sparsity assumption), and (iii) $A$ is well-conditioned, i.e., the condition factor $\kappa = \frac{\lambda_{\rm max}}{\lambda_{\rm min}}$ between the largest eigenvalue $\lambda_{\rm max}$ and the smallest non-zero eigenvalue $\lambda_{\rm min}$ is close to 1 (small condition factor assumption). 
Under these assumptions, the HHL algorithm achieves a time complexity of  $
\tilde{O}(s^2 \kappa^2  \log(d)^2 \epsilon^{-1})$, where $\epsilon$ is the error rate. This results in an exponential speed-up with respect to the system size $d$ compared to the best known classical algorithm, conjugate gradient, with runtime of $\tilde{O}( s \sqrt{\kappa} d \log(\epsilon^{-1}))$.  

The HHL algorithm serves as the quantum analogue of classical linear solvers and forms a core component of many QML applications~\cite{duan2020survey} such as  quantum least squares~\cite{chakraborty2018power, wiebe2012quantum, kerenidis2020quantum, gilyen2022improved}, kernel ridge regression~\cite{liu2017fast} and quantum $k$-means clustering~\cite{lloyd2013quantum}. 
Beyond the original framework, numerous extensions have relaxed the assumptions of HHL to improve efficiency and applicability~\cite{liu2022survey, xu2021variational,jennings2023randomized}. These include methods for ill-conditioned or dense matrices~\cite{kerenidis2020quantum, ambainis2012variable, clader2013preconditioned, tong2021fast, wossnig2018quantum}, and enhanced techniques for state preparation and Hamiltonian simulation~\cite{chakraborty2018power, kerenidis2020quantum, gilyen2022improved, low2019hamiltonian,subacsi2019quantum}.
Recent approaches based on variational algorithms have been propose to make HHL-like solvers better suited for NISQ  devices~\cite{bravo2020variational,endo2020variational, huang2019near, kyriienko2021solving,o2022near}, although these are still susceptible to the all the issues of variational QML.

It is important to note that the final output of a QLSP solvers is not the classical solution vector $\boldsymbol{x}$ itself. Instead, the algorithms produce a quantum state $\ket{\boldsymbol{x}}$ that represents the encoded form of the solution in $\left\lceil \log (d) \right\rceil$ qubits, as shown in Eq.~\eqref{eq:HHL_eq}. To recover the classical vector $\boldsymbol{x}$ from this quantum state, one must perform a measurement on the output state, and therefore, the efficiency of the HHL algorithm for classical solution depends critically on the measurement strategy~\cite{somma2021complexity}. Therefore, the HHL algorithm is most useful when it is employed as a template within larger quantum (linear-algebraic) algorithms, where a full reconstruction of the classical vector is not required~\cite{aaronson2015read}.

\subsection{Quantum spectral analysis: qPCA, QSVE, and QSVT}
\label{chapQML:subsec:qpca}
Spectral analysis plays a central role in ML, providing a means to extract the most informative features from data and to reduce its dimensionality for more efficient representation and manipulation.
Quantum algorithms build upon this idea by exploiting the inherently linear and unitary nature of quantum mechanics to perform such analyses more efficiently, particularly for high-dimensional datasets where classical methods become computationally prohibitive.

A prominent examples is the quantum PCA (qPCA) algorithm~\cite{lloyd2014quantum}, a quantum extension of the classical PCA by reformulating the classical covariance matrix as a density operator (see Sec.~\ref{chapQML:subsec:toy}). 
In qPCA, each classical data vector $ x_i$ is transformed into the normalized quantum state $\ket{ x_i}$, allowing one to define the density (covariance) operator   $\rho = \mathbb{E}[\ket{ x}\!\bra{ x}] \approx \frac{1}{n} \sum_{i=1}^n \ket{ x_i}\!\bra{ x_i}$.
Assuming access to $N$ copies of $\rho$, one can simulate its time evolution $e^{i\theta \rho}$ via density-matrix exponentiation~\cite{lloyd2014quantum}. 
This is achieved using controlled SWAP operations between copies of $\rho$ and an auxiliary state $\sigma$, such that
\[
\mathrm{Tr}_{1}\!\left[e^{-i S \Delta \theta} (\rho \otimes \sigma) e^{i S \Delta \theta}\right] = \sigma - i\,\Delta \theta [\rho, \sigma] + \mathcal{O}(\Delta \theta^2),
\]
where $\mathrm{Tr}_1[\cdot]$ denotes the partial trace over the first variable, $S$ the SWAP operator and $\Delta\theta$ a small time step. 
By repeatedly computing the trace for $N$ copies of $\rho$, the unitary operator is approximated with an error that scales as $\epsilon = 1/N$. Finally, applying Quantum QPE to this unitary $e^{i \theta \rho}$ enables extracting the spectral information of $\rho$. 
Concretely, qPCA can~\cite{wang2024comprehensive}: (i)  estimate the eigenvalues $\{\lambda_j\}$ of $\rho$, (ii) prepare eigenstates $\{\ket{u_j}\}$ with probability proportional to $\lambda_j$, and  (iii) project new input states onto the top-$k$ subspace $\sum_{j=1}^{k} \lambda_j \ket{u_j}$ by conditional postselection.  
Under the assumption that multiple copies of $\rho$ can be efficiently prepared, qPCA achieves an exponential improvement in the data dimension compared to classical PCA, with a runtime scaling as $\mathcal{O}(k \log( m))$ for constant precision $\epsilon$. \footnote{More concretely, the classical PCA can be run in time $\mathcal{O}(m^2n + m^2k)$ if $k \ll m$ due to the covariance matrix calculation and the matrix eigendecomposition, while the qPCA requires $\mathcal{O}(k\log(d))$}

An alternative approach to extracting spectral information leverages the quantum singular value estimation (QSVE) algorithm~\cite{kerenidis2016quantum}, which generalizes the principle of QPE to the singular values of a matrix. Given a matrix $A \in \mathbb{C}^{m \times n}$ we can always express it as $A= \sum_i \sigma_i \ket{u_i}\bra{v_i}$,
where $\sigma_i \ge 0$ are the singular values, and $\{\ket{u_i}\}$ and $\{\ket{v_i}\}$ are the corresponding orthonormal sets of left and right singular vectors spanning $\mathbb{C}^m$ and $\mathbb{C}^n$, respectively.  
The QSVE algorithm estimates these singular values by embedding the (generally non-unitary) matrix $A$ into a unitary operator acting on an enlarged Hilbert space. This is achieved through a block-encoding technique, which represents $A$ as the top-left block of a larger unitary operator $U_A$. Formally, one defines
\[
U_A = 
\begin{pmatrix}
    A & \cdot \\
    \cdot & \cdot
\end{pmatrix}\;, 
~~~~~A= \tilde{\Pi} U_A \Pi,
\]
where $\Pi$ and $\tilde{\Pi}$ are orthogonal projectors that isolate the subspace associated with $A$.  
Once this block-encoding is available, the QPE algorithm is applied to $U_A$ to extract the eigenphases associated with the singular values of $A$. 

Among different proposals, the QSVE work of Ref.~\cite{kerenidis2016quantum} uses a quantum walk-based algorithm, which can be also used for eigenvalue estimation~\cite{childs2010relationship}.
Starting from an input state expanded in the singular value basis, $\ket{{x}} = \sum_i\alpha_i \ket{v_i}$, the QSVE procedure transforms it into the state $\sum_i \alpha_i \ket{v_i} \ket{\bar{\sigma}_i}$ where $\bar{\sigma}_i$ denotes an estimate of the true singular value $\sigma_i$, 
satisfying $\lvert \bar{\sigma}_i \rvert = \lvert \sigma\rvert \pm \epsilon \left\Vert A \right\Vert_F$ 
with $\left\Vert\cdot\right\Vert_F$ the Frobenius norm of $A$ and $\epsilon$ the precision parameter. By measuring the ancillary register  $\ket{\bar{\sigma}_i}$, the singular values can then be retrieved. The overall runtime of QSVE scales as $
\mathcal{O}\!\left(\mathrm{polylog}(mn)/\epsilon\right)$, 
assuming that a block-encoding of $A$ and efficient state preparation oracles are available. This runtime offers an exponential improvement over classical singular value decomposition methods in the data dimension, under the same data-access assumptions.

The quantum singular value transformation (QSVT) algorithm extends the QSVE framework, allowing controlled manipulation of the matrix spectrum~\cite{kerenidis2016quantum,gilyen2019quantum, martyn2021grand}. Given a bounded polynomial or analytic function $f(x)$, QSVT algorithm constructs a sequence of unitary operations built upon quantum signal processing  tools~\cite{martyn2021grand,low2016methodology} that implement a block-encoding of the matrix function $f(A)$ directly into the quantum circuit without any intermediate measurement. In the special case where $A$ is Hermitian, the algorithm is equivalent of transforming the eigenvalues by block-encoding a matrix of the form $\sum_i f(\sigma_i) \ket{u_i}\bra{u_i}$. 
This framework generalizes many quantum linear-algebraic algorithms, including matrix inversion, projection, and spectral filtering, within a single unified approach. It also achieves optimal asymptotic scaling with respect to both the precision $\epsilon$ and condition number $\kappa$, and supports theoretical speedups that range from quadratic to exponential~\cite{gilyen2019quantum}.
By appropriately choosing the transformation function $f(x)$, QSVT serves as a fundamental subroutine in a wide range of QML applications~\cite{martyn2021grand}, including Hamiltonian simulation~\cite{gilyen2019quantum,low2017optimal}, quantum linear system solving~\cite{wossnig2018quantum,gilyen2019quantum, orsucci2021solving}, quantum recommendation system~\cite{kerenidis2016quantum}, and topological data analysis~\cite{lloyd2016quantum}.

In the next section, we provide a concrete example of QML application, the quantum recommendation system, which employs the QSVE/QSVT algorithm as a subroutine for a efficient low-rank matrix reconstruction.

\subsection{Application: Quantum recommendation system}

In our everyday life, recommendation systems are ubiquitous, providing personalized suggestions based on a user's past behaviors and preferences on social media, e-commerce, and streaming platforms such as Netflix, Amazon, and YouTube~\cite{ricci2021recommender, roy2022systematic}. These systems aim to predict which products a user is likely to appreciate, given partial historical information about user–product interactions.  Due to their practical importance, many ML techniques have been proposed to tackle this problem, such as kNN methods, deep learning, graph-based models, reinforcement learning, and even large language models~\cite{raza2024comprehensive}. 

In the box below, we present he basic mathematical framework for recommendation system.

\begin{mybox}{Mathematical formulation for recommendation system} 

Let $P \in\mathbb{R}^{m \times n}$ be the preference matrix, such that $P_{ij}$ represents how the user $i$ values the product $j$. Typically, $m$ (number of users) and $n$ (number of products) are very large, on the order of $10^6 - 10^8$, prohibiting the storage or heavy computation of the matrix. To simplify the analysis, we can round $P$ into a binary preference matrix $T \in \mathbb{R}^{m \times n}$, where,  
\[
T_{ij} = \begin{cases}
    1 & \text{if user $i$ likes product $j$} \\
    0 & \text{otherwise}\;. 
\end{cases}
\]

In practice, most entries in this matrix are missing because the system receives data in an online manner. In other words, not all elements of the preference matrix are known in advance -- the system only obtains information and updates the entry $P_{ij}$, when a user takes an action, such as writing a review or purchasing a product. The resulting data is, therefore, incomplete and highly sparse. The goal of the recommendation system is to recommend suitable products to user $i$, among the products with missing entries that the user has not seen yet. To address this sparsity, a uniformly subsampled matrix $\hat{T}$ is constructed from the observed entries of $T$, by assigning each known entry $\hat{T}_{ij} = T_{ij}/p$  with a probability $p$, and zero, otherwise.  

The key assumption in classical recommendation systems is that the underlying preference matrix $T$ is approximately low-rank, meaning it can be well-approximated by a matrix of rank $k$, where $k \ll \min(m, n)$. This assumption has both ``philosophical'' and empirical justification: user preferences typically depend on a small number of latent factors (e.g., genres, styles, or categories) independent on the number of the users $m$ or the number of products $n$, which induce patterns of correlated interests~\cite{kerenidis2016quantum}. Thus, most of the information in $T$ can be captured by a low-dimensional subspace and the users in general have tendency to value a few products than to spread their preferences across all possible items.   
\label{subsec:qrec}
\end{mybox}

Classical recommendation system consists of three steps: (i) subsampling, i.e., constructing $\hat{T}$, (ii) low-rank approximation or matrix reconstruction: we construct a low-rank approximation $\hat{T}_k$ of the subsampled matrix $\hat{T}$ using a singular value algorithm, or other factorization techniques, and (ii) prediction or sampling: for a given user $i$, we estimate the good recommendations by projecting $T_k$ on the user's feature vector. Such an algorithm takes a runtime of $\mathcal{O}(\textrm{poly}(mnk) + \textrm{poly}(nk))$, with each term corresponds to the complexity of low-rank approximation and sampling for per-user recommendation, respectively. With the increasing amount of information to be treated, the values of $m$ and $n$ become large to be efficiently managed by the classical algorithm.

The previous steps can be reformulated in the a quantum setting by leveraging quantum linear-algebraic subroutines, such as qPCA, QSVE and QSVT (see Sec.~\ref{chapQML:subsec:qpca}). Indeed, the quantum recommendation system (QRS)~\cite{kerenidis2016quantum} algorithm can achieve a runtime scaling as $\mathcal{O}(\text{poly}(k)) \text{polylog}(mn))$ under the assumption of efficient quantum data access, and thus, has been believed to exhibit an exponential speed-up over the classical algorithm. The main assumption in the algorithm to ensure the potential speed up is that the data is stored in a classical data structure, for instance the binary tree explained above, which enables the quantum algorithm to efficiently create superposition of rows of the subsample matrix within $\mathcal{O}(\mathrm{polylog}(mn))$ time and memory. In the box below, we provide the detailed description of the QRS algorithm.

\begin{mybox}{Quantum Recommendation System} 

Here, we summarize the detailed procedure of QRS presented in Ref.~\cite{kerenidis2016quantum}, leveraging quantum linear-algebraic primitives presented in the last sections to perform an efficient low-rank approximation of the given preference matrix, $T \in \mathbb{R}^{m\times n}$.

The first step is to perform pre-processing.  As in classical recommendation systems, the procedure begins by subsampling the preference matrix $T$ to obtain a matrix $\hat{T}$ with entries $\hat{T}_{ij} = T_{ij}/p$  with a probability $p$, and zero, otherwise.      Each row $\hat{T}_i$ is stored in the quantum-accessible data structure, such as the binary tree explained above. Under the assumption of an efficient quantum data structure, this step can be performed in $\mathcal{O}(\mathrm{polylog}(mn))$. 
    
Next, we initialize the quantum computer to the state $\ket{\hat{T}_i}$, corresponding to row for user $i$, which is prepared by loading it from the data structure. This normalized state represents a superposition over all products weighted by their known preference values, serving as the input for the subsequent projection step. 

From here, we reach the core of the QRS algorithms, a quantum projection with threshold, where one projects the user’s preference vector onto the subspace spanned by the dominant singular vectors of $\hat{T}$.  Unlike classical methods, which explicitly reconstruct a low-rank approximation of $T$, the quantum setting uses qPCA or QSVE algorithms to estimate the singular values $\{\sigma_i\}$ and singular vectors $\{\ket{u_i}, \ket{v_i}\}$ of $\hat{T}$. Here, instead of selecting a fixed number of top-$k$ singular values, the algorithm retains those within a range $[(1-\tau)\sigma,\, \sigma) \bigcup [\sigma, \sigma_{\rm max}]$, where $\sigma_{\rm max}$ is the maximum  singular value, and $\sigma, \tau$ the parameters to control the precision of the reconstruction.  Given the decomposition $\hat{T} = \sum_i \sigma_i \ket{u_i}\!\bra{v_i}$, the thresholded approximation matrix is defined as 
    \[
    \hat{T}_{\sigma, \tau} = \sum_{ \sigma_j\in [(1-\tau) \sigma, \sigma) \cup [\sigma, \sigma_{\rm max}] } \sigma_j \ket{u_j}\!\bra{v_j}\;.
    \] 
As shown in Ref.~\cite{kerenidis2016quantum}, the approximated matrix $\hat{T}_{\sigma, \tau}$ satisfies the bound $\|T - \hat{T}_{\sigma, \tau} \|_F  \le 9\epsilon \|T\|_F$ with high probability, when the parameters are chosen as $\sigma = \epsilon \sqrt{p} \|\hat{T}\|_F/\sqrt{2k}$ and $\tau = 1/3$ for $k$ satisfying that $\| T - T_k\|_F \le \epsilon\|T\|_F$. This ensures that $\hat{T}_{\sigma,\tau}$ is a high-quality approximation to the original matrix $T$ without the need for explicit reconstruction. To obtain the predicted preference vector for user $i$, the QSVE-based quantum projection algorithm applies this spectral filtering to the user’s state $\ket{\hat{T}}_i$, returning the output projected vector state $\ket{\left( \hat{T}_{\sigma, \tau}\right)_i} = \ket{\hat{T}_{\sigma, \tau}^{+} \hat{T}_{\sigma, \tau}\hat{T}_i}\;$ where $\hat{T}_{\sigma, \tau}^{+}$ denotes the pseudoinverse of $\hat{T}_{\sigma, \tau}$. This resulting state represents the projection of user $i$’s preferences into the low-dimensional space spanned by the singular vectors that encodes the most informative features. The procedure succeeds with probability at least $1 - 1/\mathrm{poly}(n)$ and has an expected runtime of $\mathcal{O}(\mathrm{poly}(k)\mathrm{polylog}(mn))$ with an additional scaling depending on the Frobenius norm of $T$ and the projected matrix $\hat{T}_{\sigma, \tau}$.  

Finally, the state $|\left(\hat{T}_{\sigma, \tau}\right)_i\rangle$ is measured in the computational basis with probability $|(\hat{T}_{\sigma, \tau})_{ij}|^2/\|\left(\hat{T}_{\sigma, \tau}\right)_i\|^2$. The outcome $j$ with the highest probability corresponds to the product that the user $i$ is most likely to appreciate. 

\end{mybox}

It is important to note that, similar to the HHL algorithm, the QRS produces a quantum state representing the final projected vector rather than an explicit classical output. However, unlike the HHL algorithm, which aims to recover the exact numerical solution to a system of linear equations, the QRS does not require obtaining the full row vector of the reconstructed matrix. 
Instead, it is sufficient to sample from this quantum state to identify the output with the highest probability, which directly corresponds to the most relevant recommendation.

\subsection{On dequantizations and data-access assumptions \label{subsec:dequantization-bridge}}

The era of 2010 was perhaps the golden era for QML. Algorithms, such as the QRS, based on algebraic primitives had lead to exponential advantages for practical tasks with high relevance for industry. As such, one may wonder: \textit{How can we reconcile this result with the law of conservation of weirdness? How can an exponential advantage be reached for an unweird task such as recommendation systems which Netflix uses every day?}

As everything, the devil is in the details. In the earlier discussion of QRS, the reported quantum speedup relies heavily on a strong assumption that the algorithm can efficiently access the data structure, requiring only a polylogarithmic number of queries to the input data~\cite{kerenidis2016quantum,aaronson2015read}. However, this assumption is far from trivial~\cite{jaques2023qram, arunachalam2015robustness}. Indeed, in Ref.~\cite{tang2019quantum} it was proven that, contrary to the earlier belief that the QRS achieves an exponential speedup over the best-known classical algorithms, the improvement is in fact at most polynomial when compared to a carefully designed quantum-inspired classical algorithm. The proposed method employs a  classical data structure analogous to that used in the quantum model by reformulating the notion of quantum superposition within a classical setting, specifically by replacing the state preparation assumption with an $\ell_2$-norm sampling/query access assumption. This breakthrough introduced the idea of dequantization, which challenges the existence of all exponential quantum advantages developed in the mid 2010's. The concept of dequantization has since provided valuable insight into the role of data access models in QML, motivating researchers to reexamine earlier claims of quantum advantage in the settings. For further discussion on this topic, readers may refer to Refs.~\cite{tang2019quantum, tang2021quantum, tang2022dequantizing, cotler2021revisiting}.

Still, despite these dequantizations, the hope still remains for polynomial improvements when comparing classical and quantum algorithms. Moreover, the key idea of using basic quantum algorithms as building blocks in larger data processing schemes remains and interesting and promising research avenue, with some new proposals studying avenues for obtain large advantages without the need of components such as QRAMs~\cite{harrow2020small}.

\section{Other forms of quantum machine learning}
In the previous sections we have focused on QML for supervised learning, as a significant amount of effort has been put forward toward understanding how and where quantum models can enhance this form of learning. This section broadens our score, as we delve into  three additional pillars of QML research: quantum generative models, quantum unsupervised learning, and quantum reinforcement learning.  

\subsection{Quantum generative models}
\label{chapQML:subsec:QGM}

Quantum computers are, by definition, probabilistic. The fact that quantum states can exist as a superposition of the elements in a basis $\{\ket{x_i}\}_i$, i.e., $\ket{\psi}=\sum_ic_i |x_i\rangle$ with $c_i \in \mathbb{C}$, implies that measurements on said basis yield a probabilistic outcome. Specifically, one obtains the result $x_i$ with probability $|c_i|^2$ as per Born's rule. Indeed, this feature is exploited in quantum supremacy experiments, where a quantum computer solves a task that no  classical computers can: sampling probabilities at the output of random quantum circuits~\cite{aaronson2016complexity,boixo2018characterizing,google2019supremacy,kondo2022quantum}. While such experiments serve as mathematical demonstrations (with no known usefulness), they provide the basic mathematical formalism for quantum generative models, which seeks to leverage the intrinsic probabilistic nature of quantum computers to tackle practical learning tasks based on probability distributions.

Earlier, in Sec.~\ref{chapQML:subsec:PAC}, we introduced the notion of PAC learning for Boolean function classes, which provides a theoretical foundation for supervised learning. Notably, the PAC framework can also be extended to the context of generative modeling, where the goal is to approximately learn an unknown distribution from samples~\cite{ coyle2020born, sweke2021on,kearns1994learnability}. Specifically, consider the set $\mathbb{D}_n = \{\mathcal{D} \; | \; \mathcal{D}:\{0, 1\}^n \to [0, 1]\}$, the class of discrete distributions over $\{0, 1\}^n$ for an integer $n \ge 1$, and $\mathcal{C} \subseteq \mathbb{D}_n$ a distribution (or concept) class.
To analyze the computational complexity of learning distributions, it is useful to distinguish between efficient generators and efficient evaluators. A class $\mathcal{C}$  is said to be efficiently generable if, for any distribution $\mathcal{D} \in \mathcal{C}$, there exists a circuit\footnote{Here, the term circuit refers not only to a quantum circuit but also to a classical logical circuit, as the concept of learnability of discrete distributions was originally developed for classical distribution~\cite{kearns1994learnability}.} $G_\mathcal{D}$ which can generate samples in $\{0, 1\}^n$ according to the distribution $\mathcal{D}$ within polynomially-sized resources. On the other hand, the class $\mathcal{C}$ is said to be efficiently evaluable if for any distribution $\mathcal{D} \in \mathcal{C}$, there exists a circuit $E_\mathcal{D}$ which can output the probability density $\mathcal{D}(x)$ for any input $x \in \{0, 1\}^n$ within the polynomial resource bounds. This distinction separates models that are learnable to generate from those that are learnable to evaluate. In the literature, evaluators correspond to explicit models, which define the probability density explicitly, such as Variational Autoencoders (VAEs)~\cite{kingma2019introduction} or Energy-Based Models (EBMs). Generators, on the other hand, correspond to implicit models, which define the distribution only through a sampling process—such as Generative Adversarial Networks (GANs)~\cite{goodfellow2014generative}.

To qualify how closely a learned distribution $\hat{\mathcal{D}}$ approximates the target distribution $\mathcal{D}$, one typically uses compares distributions using metrics such as the Kullback--Leibler (KL) divergence\footnote{When dealing with implicit models, it is more convinient to use metrics such as the maximum mean discrepancy (MMD)~\cite{rudolph2023trainability, gretton2012kernel}.} 
\[
KL(\mathcal{D}\|\hat{\mathcal{D}}) = \sum_{x \in \{0, 1\}^n} \mathcal{D}(x) \log \left(\frac{\mathcal{D}(x)}{\hat{\mathcal{D}}(x)}\right)\;.
\]
The KL divergence is non-negative and equals zero if and only if the two distributions coincide exactly, i.e., $\mathcal{D}(x) = \hat{\mathcal{D}}(x)$ for all $x$. 
Although it is not a true metric as it is asymmetric and does not satisfy the triangle inequality, it serves as a natural measure of how much information is lost when $\hat{\mathcal{D}}$ is used to approximate $\mathcal{D}$. Moreover, it upper bounds several other common distance measures, such as $L_1$ norm. This leads to the definition $\epsilon$-good generator and $\epsilon$-good evaluator.
A generator $G_\mathcal{D}$ (or evaluator $E_\mathcal{D}$) is called $\epsilon$-good if the distribution $\mathcal{D}$ it produces satisfies $KL(\mathcal{D}\|\hat{\mathcal{D}}) \le \epsilon$.  
This definition mirrors the notion of PAC learning, where the goal is to approximate an unknown target distribution within a specified accuracy $\epsilon$.  

From the previous, we can define the notion of efficiently PAC learnable for distribution learning. 
Given access to a sampling oracle $O(\mathcal{D})$, a distribution class $\mathcal{C}$ is said to be efficiently $\epsilon$-PAC-learnable with a generator (or an evaluator) if a learning algorithm can produce an $\epsilon$-good generator (respectively, evaluator) for all $\mathcal{D} \in \mathcal{C}$ using polynomial resources.  
Importantly, a class that is efficiently PAC-learnable with a generator may not be efficiently PAC-learnable with an evaluator~\cite{kearns1994learnability}.  
Learning by generation is often easier, whereas learning by evaluation can be computationally intractable, often leading to \#P-hard problems.  

Now, one can ask questions such as~\cite{sweke2021on}: \emph{Does there exist a class of probability distributions that (i) is efficiently classically generable, (ii) is \textbf{not} classically PAC-learnable, yet (iii) is quantum PAC-learnable?}\footnote{This study on learnability is constrained on learning the distribution defined by underlying classical processes, and under the assumption that we have access only to the classical sample oracle to learn classical generators. However, similar discussion is also possible for quantum samples (see Section V in Ref.~\cite{sweke2021on}).} 
To simplify the analysis, most studies focus on the learnability of generators.\footnote{Although this section we only discusses the separation between quantum-classical learnability with generators, similar arguments can be made with evaluators~\cite{pirnay2023superpolynomial}. 
} 
A canonical example is provided by the class of pseudo-random functions (PRFs), which are provably not classically PAC-learnable even with efficient generators~\cite{kearns1994learnability}.  

The classical hardness of learning PRFs relies on standard cryptographic hardness assumptions, specifically the decisional Diffie--Hellman (DDH) assumption~\cite{boneh1998decision}. 
For a cyclic group $G$ of prime order $q$ with a group generator $g$, the DDH assumption asserts that no classical probabilistic polynomial-time algorithm can distinguish tuples of the form $(g^a, g^b, g^{ab})$ from $(g^a, g^b, g^c)$, where $a,b,c \in \mathbb{Z}_q$. 
This assumption underpins many cryptographic protocols, as it guarantees the classical hardness of distinguishing PRFs from truly random functions. 
However, in the quantum setting, this assumption no longer holds as quantum algorithms such as Shor’s algorithm  can efficiently compute discrete logarithms and thus distinguish such tuples in polynomial time.  
Under this separation of computational assumptions, it has been proven that there exists a discrete distribution class $\mathcal{C} \subseteq \mathbb{D}_n$ which is provably not efficiently PAC-learnable by any classical learning algorithm, yet is efficiently PAC-learnable by a quantum learning algorithm~\cite{sweke2021on}.  

This theoretical separation for these contrived distributions, motivates the design of quantum generative models (QGMs) to learn complex probability distributions that may be unreachable by classical generative models but also useful in practice. 
Quantum circuits can naturally act as probabilistic discrete data generators, giving rise to architectures such as quantum Boltzmann machines (QBMs)~\cite{Amin2018Quantum, coopmans2023sample}, quantum circuit Born machines (QCBMs)~\cite{rudolph2023trainability,liu2018differentiable, kyriienko2024protocols}, quantum variational autoencoders (qVAEs)~\cite{khoshaman2018quantum, li2022scalable, gao2020high},  quantum generative adversarial networks (qGANs)~\cite{zoufal2019quantum, zoufal2021generative, huang2021quantumgenerative, bermot2023quantum, tsang2023hybrid}, and having been used across a wide range of domains~\cite{alcazar2020classical,zhu2022generative, kondratyev2021non, assouel2022quantum, zeng2019learning,di2024quantum, kiss2022conditional, delgado2022quantum,kieferova2021quantum,kieferova2021quantum}. In parallel, the quantum generative models are also proposed as a quantum state generators~\cite{patel2024quantum} (quantum state learning~\cite{benedetti2019generative}), or state initialization~\cite{zoufal2019quantum}. 
Beyond discrete distribution learning, quantum generative models have also been extended to continuous data~\cite{romero2021variational, bravo2022style, chang2024latent, barthe2025parameterized}, where the output distribution corresponds to a continuous sample generated by injecting noise drawn from a prior distribution.  

Despite significant recent advances~\cite{coyle2020born, huang2021quantum, huang2025generative}, most existing QGMs remain heuristic and empirical, with limited theoretical understanding or formal performance guarantees compared to quantum supervised learning. 
Bridging this gap between empirical success and theoretical foundations remains one of the most active research frontiers in QML~\cite{huang2025generative, gili2024generalization}.

\begin{mybox}{Examples of quantum generative models}

\textbf{Quantum Boltzmann machine (QBM)}
A foundational model in quantum generative learning is the QBM~\cite{Amin2018Quantum}, which generalizes the classical Boltzmann Machine (BM)~\cite{ackley1985learning, hinton1986learning} to the quantum regime.  
The classical Boltzmann machine consists of an array of visible units $\boldsymbol{v}$, corresponding to observed variables (inputs or outputs), and an array of hidden units $\boldsymbol{h}$, which capture latent correlations and provide additional representational capacity.  

Using the joint variable $\boldsymbol{z} = \{\boldsymbol{v}, \boldsymbol{h}\}$, the probability of observing a certain configuration $\boldsymbol{v}$ of the output visible unit is defined as
\[
p_{\boldsymbol{\theta}}(\boldsymbol{v}) = \sum_{\boldsymbol{h}} \frac{e^{-E(\boldsymbol{z})}}{\mathcal{Z}}, 
\quad 
E(\boldsymbol{z}) = \sum_i \theta_i z_i + \sum_{i,j} \theta_{ij} z_i z_j,
\]
where $\mathcal{Z} = \sum_{\boldsymbol{v}, \boldsymbol{h}} e^{-E(\boldsymbol{v}, \boldsymbol{h})}$ is the partition function ensuring normalization.  
The trainable parameters $\boldsymbol{\theta}$ are optimized by minimizing the system’s energy or equivalently maximizing the likelihood of observed data.  
Since this optimization process closely parallels the search for a system’s ground state, quantum mechanics provides a natural extension—replacing classical binary units with qubits and energies with Hamiltonians.

Already proven to be a universal function approximator, the classical BM can represent highly complex joint probability distributions.  
In the quantum setting, the target data distribution is represented by a density matrix $\eta$.  
For a quantum system with a Hamiltonian $H(\boldsymbol{w})$ parametrized by a real values $\boldsymbol{w}$, the density matrix of the system is given by 
\[
\rho(\boldsymbol{w}) = \frac{e^{-H(\boldsymbol{w})}}{\mathcal{Z}},~~~~\mathcal{Z}(\boldsymbol{w}) = \mathrm{Tr}[e^{-H(\boldsymbol{w})}]\;.
\] 
Given a state $\ket{\boldsymbol{v}}$ corresponding to a visible configuration, its probability is obtained from the reduced density matrix as
\begin{equation}
p(\boldsymbol{v}) = \mathrm{Tr}\!\left[\Lambda_{\boldsymbol{v}} \rho \right],
\quad 
\Lambda_{\boldsymbol{v}} = \ket{\boldsymbol{v}}\!\bra{\boldsymbol{v}} \otimes I_{\boldsymbol{h}}.
\end{equation}

A commonly used Hamiltonian in QBM implementations is the transverse-field Ising model, $ H = -\sum_i \Gamma_i X_i - \sum_i b_i Z_i - \sum_{i<j} w_{ij} Z_i Z_j$,
where $\Gamma_i$ are transverse-field coefficients, $b_i$ are bias terms, and $w_{ij}$ are coupling strengths between qubits.  
The parameter set $\boldsymbol{\theta} = \{\Gamma_i, b_i, w_{ij}\}$ defines the learnable parameters of the model.
During the training~\cite{patel2024natural}, the parameters $\boldsymbol{\theta}$ are updated to minimize either the negative log-likelihood
\[
\mathcal{L} = -\sum_{\boldsymbol{v}} p^{\eta}(\boldsymbol{v}) 
\log \left(\frac{\mathrm{Tr}\!\left[\Lambda_{\boldsymbol{v}} e^{-\beta H}\right]}{\mathrm{Tr}\!\left[e^{-\beta H}\right]}\right),
\]
where $p^{\eta}(\boldsymbol{v}) = \mathrm{Tr}\!\left[\Lambda_{\boldsymbol{v}} \eta \right]$ represents the target data distribution, or alternatively quantum relative entropy (KL divergence) $
\mathcal{L} = \mathrm{Tr}\!\left[\eta \log (\eta) - \eta \log (\rho) \right]$.

The major computational challenge for Boltzmann machines arises from the need to sample from thermal (Gibbs) distributions.  
Classically, computing the partition function $\mathcal{Z}$ of the Ising model is \#P-hard~\cite{bremner2016average}, rendering exact thermal-state preparation intractable~\cite{kieferova2017tomography}.  
As a result, most of the classical algorithms rely approximate methods such as contrastive divergence~\cite{hinton2002training}.  
Similarly, exact QBM training is BQP hard and intractable on the near term quantum hardware as it requires quantum thermal state preparation of non-commuting Hamiltonians.
To address this issue, restricted forms of QBMs have been proposed~\cite{Amin2018Quantum}, where the Hamiltonian includes only commuting terms (for example, by omitting the transverse-field term).  
This simplification allows efficient simulation on near term hardware using shallow circuits but eliminates quantum correlations, thereby limiting the possibility of quantum advantage.  
Alternative approaches include hybrid and variational training schemes, such as the variational QBM~\cite{zoufal2021variational}, the evolved QBM~\cite{minervini2025evolved} and tomography-based training methods~\cite{kieferova2021quantum,kieferova2017tomography, xiao2020quantum}, which approximate Gibbs-state sampling through optimization.

\vspace{0.4cm}
\textbf{Quantum Circuit Born Machine (QCBM)} 
Another major class of quantum generative models is that of QCBMs~\cite{coyle2020born,liu2018differentiable}, which generates samples according to the Born rule, exploiting the intrinsic probabilistic nature of quantum mechanics.  
A QCBM defines a probability distribution over $n$-bit strings $x \in \{0,1\}^n$ as $p_{\boldsymbol{\theta}}(x) = \left| \langle x | \psi(\boldsymbol{\theta}) \rangle \right|^2 
= \left| \langle x | U(\boldsymbol{\theta}) | 0 \rangle^{\otimes n} \right|^2$, where $U(\boldsymbol{\theta})$ is a parameterized quantum circuit acting on $n$ qubits, and $\ket{0}^{\otimes n}$ is the initial state.

Given a dataset $\mathcal{X} = \{x\}$ (not necessarily discrete) sampled from an unknown target distribution $\mathcal{D}$, the goal is to learn parameters $\boldsymbol{\theta}$ that produce a model distribution $\hat{\mathcal{D}} = p_{\boldsymbol{\theta}}$ approximating $\mathcal{D}$.  
The parameters are trained by minimizing a divergence-based loss, such as the KL divergence or the maximum mean discrepancy~\cite{rudolph2023trainability}.  
When the data are continuous, $\mathcal{X}$ is typically discretized into finite bins to map the data distribution on the discrete outcome space. 

Unlike QBM, the QCBM typically requires only a shallow depth circuit, and thus serves as an efficient, tractable quantum generator on near term quantum hardware~\cite{coyle2020born}.  The original QCBM employed a hardware efficient ansatz; however, several improved variants have been proposed to enhance expressivity and trainability, including tensor-network-inspired architectures~\cite{du2020expressive} and Hamiltonian-variational ans\"atze~\cite{coyle2020born}.  Still more work is needed to understand how QCBMs can overcome the well known limitations of variational QML techniques. For further developments and recent progress in QCBMs, readers are referred to Refs.~\cite{du2020expressive,zhu2019training, monaco2022quantum, nietner2025average}.

\end{mybox}

\subsection{Quantum unsupervised learning }
While generative models aim to learn the underlying probability distribution of a dataset in order to generate new, unseen samples, other forms of unsupervised learning focus on identifying underlying structural correlations and relationships within the data~\cite{kerenidis2019q, gujju2024quantum}. Although research on quantum unsupervised learning beyond generative models is still in its early stages, several interesting approaches have been proposed. However, theoretical investigations in this area, including the learnability gap between classical and quantum unsupervised learning, remain limited and as the result, this section is primarily explanatory rather than analytical compared to the preceding ones.

To begin, let us discuss the quantum autoencoder (QAE)~\cite{wan2017quantum,romero2017quantum, pepper2019experimental, cao2021noise, ma2024quantum, bravo2021quantum}, which serves as the quantum extension of the classical autoencoder for data compression~\cite{rumelhart1986parallel}.  
Classically, an autoencoder consists of two distinct neural networks, an encoder that maps high-dimensional input data to a low-dimensional latent representation, and a decoder that reconstructs the input from this latent space.  This provides a compact encoding of the data, leading to more efficient computation and storage while revealing meaningful underlying structure in the representation space.
In the quantum setting, the QAE performs an analogous task for a collection of $n$-qubit quantum states, $ {\ket{\psi^{\mathrm{(in)}}_i}}$, defined on a bipartite system $A + B$ composed of subsystems $A$ and $B$. The QAE learns to compress these $n$-qubit pure states into a set of compressed states ${\ket{\psi_i^{\mathrm{(comp)}}}}$ defined on a smaller subsystem $A$ of $m$ qubits, where $m < n$. The goal is to preserve the essential quantum information while minimizing the information loss in the discarded subsystem $B$, thereby maximizing the similarity between the original state and its reconstruction on the composite system $A + B'$. This is often achieved by training a PQC, and minimizing loss functions based on  local observables~\cite{bravo2021quantum}, quantum mutual information for mixed input states~\cite{ma2024quantum}, or the occupation probability of discarded modes in photonic implementations~\cite{pepper2019experimental}. The QAE framework can also be adapted for classical data by encoding it into a quantum Hamiltonian and performing compression on the thermal states of the Hamiltonian~\cite{cao2021noise}. In practice, quantum autoencoders are often used as a method to prepare low-dimensional quantum input states for subsequent quantum algorithms~\cite{zhang2021generic, locher2023quantum,cao2019quantum,bondarenko2020quantumautoencoders, mok2024rigorous,ngairangbam2022anomaly}. As always, we remind the reader that variational QAEs are extremely prone to trainability barriers such as barren plateaus~\cite{larocca2024review,cerezo2020cost}, indicating that one must be extremely careful when applying them at large scales.

Another important class of quantum unsupervised learning algorithms is quantum clustering, which aims to group data samples into clusters based on their similarity~\cite{kerenidis2019q,lloyd1982least, wu2022quantum, kavitha2023quantum, magano2023quantum, diadamo2022practical}. 
Given an input dataset $\mathcal{S}$ containing $N$ samples of $d$-dimension and a predefined number of clusters $k$, the goal of clustering is to partition the dataset into $k$ groups  such that samples within the group have the maximum similarity between them than those in the other groups. At high level, quantum clustering algorithm follows the same general procedure as the classical algorithms, but replaces key subroutines with quantum subroutines. 
Under the standard assumption of an efficient quantum data access model (see Sec.~\ref{sec:chapQML:qml-linear-algebra}), one can show an asymptotic improvement in computational complexity over the classical methods~\cite{ kerenidis2019q, aimeur2013quantum, kerenidis2021quantum}. For example, the quantum $k$-means clustering (or q-means) achieves a runtime of $\mathcal{O}(k^{2.5}d \cdot \mathrm{polylog}(N))$\footnote{The exact complexity depends on additional factors as condition number of data matrix or the desired precision, but we provide a simplified expression for a general intuition on the scaling.}, reducing the dependence on the number of samples compared to the classical $k$-means with runtime of $\mathcal{O}(kdN)$.

Spectral clustering represents an alternative clustering approach that operates in the graph domain when the cluster structures are highly non-convex. By employing quantum spectral analysis primitives such as QSVE or QSVT (see Sec.~\ref{chapQML:subsec:qpca}), the algorithm exploits the spectral properties of the graph Laplacian matrix to uncover intrinsic relationships between data points.
While the classical spectral clustering algorithm requires runtime of  $\mathcal{O}(N^3)$ resulting from the eigendecomposition, the quantum spectral clustering reduces this to a linear scaling of $\mathcal{O}(N)$, offering a potential quantum advantage~\cite{kerenidis2021quantum, li2022quantum}. 
Other variants of quantum clustering include density-based clustering~\cite{magano2023quantum}, classical-quantum hybrid approach~\cite{poggiali2024quantum}, variational method ~\cite{bermejo2023variational, fang2024quantum}, or optimization-based method~\cite{otterbach2017unsupervised, matsumoto2022distance}. 
The box below presents detailed procedures for the q-means and quantum spectral clustering algorithms. 
 
Overall, quantum unsupervised learning provides a promising avenue other than the standard supervised learning. Yet, this field remains in its early stages and a comprehensive understanding of their scalability and practical advantages is still an active area of research. 

\begin{mybox}{Example of quantum clustering algorithms} 

In this box, we present the detailed steps of the quantum clustering algorithm introduced above.
Consider a classical dataset $\mathcal{S} = \{x_i\}_{i=1}^{N}$ consisting of $N$ $d$-dimensional samples.
The dataset can be represented as a data matrix $S\in \mathbb{R}^{N \times d}$, where the $i$-th row corresponds to the data vector $x_i$. The data matrix is assumed to be efficiently stored and accessed through QRAM~(see Sec.~\ref{sec:chapQML:qml-linear-algebra}).
Throughout this section, we use $\|\cdot\|_2$ to denote the Euclidean norm (or $\ell_2$ norm), and $\widetilde{\|\cdot\|}_2$ to denote its estimated value, whose error is bounded as  $| \|x_ i - x_ i  \|_2 - \widetilde{ \|x_ i - x_ i  \|}_2 | < \epsilon$ for some $\epsilon > 0$.

\vspace{0.3cm}
\textbf{Quantum $k$-means clustering}~\cite{kerenidis2019q}
Quantum $k$-means clustering, or q-means clustering, is a quantum extension of classical $k$-means clustering algorithm. In the classical setting, the algorithm begins by selecting $k$ initial centroids, each associated with a distinct cluster label. It then iteratively performs two alternating steps: (i) each data point is assigned to the cluster whose centroid is closest in Euclidean distance, and  (ii) each centroid is updated by taking the mean of all data points assigned to that cluster. The procedure repeats until convergence, either when the change in the centroids falls below a predefined threshold or when the maximum number of iterations is reached. 

In the quantum setting, the algorithm takes the same high-level procedure, but replaces key subroutines with quantum linear-algebraic primitives that can yield asymptotic speedups. The algorithm begins by choosing $k$ initial centroids $c^0_1, \cdots c^0_k$ and storing them in a QRAM structure. At each iteration $t$, the quantum algorithm prepares an initial state that encodes all data points and its estimate distance from all the cluster centroids as $\sum_{i =1}^N \ket{i}(\sum^k_{i=1} \ket{j}\ket{0}) \to \sum_{i =1}^N \ket{i}(\sum^k_{j=1} \ket{j}\ket{ \widetilde{\|x_i - c_j \|}^2_2 })$ (normalization factor simplified). A quantum minimum-finding procedure is then applied to determine, for each data point, the minimum distance $\min_j(\|x_i - c_j \|_2)$. By uncomputing the quantum states, the algorithm creates a superposition of all the data points and their labels corresponding to the centroid with a minimum distance. The resulting state can be written as $\sum_{i =1}^N \ket{i}\ket{\ell^t(v_i)} $ where $\ell^t(x_i) = \mathrm{argmin}_j \widetilde{\| x_i - c^t_j\|}_2 $. 

The next stage computes the new centroid for each cluster $\{C_j^t\}_{j=1}^k$, where $|C_j^t|$ denotes the number of samples associated with the cluster $j$. Measuring the label register yields, for each cluster, a normalized quantum state representing a superposition of all data points belonging to that cluster as $\ket{\mathcal{X}^t_j} = (|C_j^t |)^{-\frac{1}{2}} \sum_{i \in C^t_j} \ket{i}$. The centroid state $\ket{C^{t}_j}$  can be finally obtained by multiplying matrix $S^T$ to the state $\ket{\mathcal{X}^t_j}$. These updated centroids $c^{t+1}_1, \cdots c^{t+1}_k$ are then stored in QRAM.

The improvement in computational complexity of q-means over the classical $k$-means algorithm arises primarily from the efficiency of quantum subroutines for distance estimation, minium-finding, and matrix operations. A detailed theoretical analysis of these steps can be found in~\cite{kerenidis2019q}.

\vspace{0.3cm}

\textbf{Quantum Spectral Clustering}~\cite{kerenidis2021quantum}
Unlike $k$-means clustering, which directly measures Euclidean distances between data points, spectral clustering leverages the spectral properties of a similarity graph to embed the data in a low-dimensional space before clustering. This approach is particularly effective for datasets with highly non-convex cluster structures, where geometric separation is non-trivial in the original data space.  

The algorithm begins by constructing a similarity graph in which each vertex represents a data point. Two vertices $i$ and $j$ are connected if $\|x_i - x_j\|_2 \le d_{\min}$, where $d_{\min}$ is a predefined threshold. This graph can be represented by a similarity matrix $A = \{a_{ij}\} \in \mathbb{R}^{N \times N}$, where $a_{ij} = 1$ if the vertices $i \ne j$ are connected and $a_{ij} = 0$ otherwise. From $A$, an incidence matrix $B \in \mathbb{R}^{N \times \frac{N(N+1)}{2}}$ is constructed, where each row $i$ represents a vertex and each column $(p,q)$ a possible edge between vertices $p$ and $q$. The matrix elements are defined as $b_{i,(p,q)} = a_{pq}$ if $i = p$, $b_{i,(p,q)} = -a_{pq}$ if $i = q$, and $b_{i,(p,q)} = 0$ otherwise. 

The normalized Laplacian matrix is then given by $\mathcal{L} = \mathcal{B} \mathcal{B}^T \in \mathbb{R}^{N \times N}$ with $\mathcal{B}$ the normalized incidence matrix. As the normalized Laplacian matrix is symmetric and positive-semidefinite, its eigenvalues are real and non-negative. Mathematically, the eigenvalues and corresponding eigenvectors of $\mathcal{L}$ capture the intrinsic connectivity and structural properties of the similarity graph. The first $k$ eigenvectors with smallest eigenvalues, representing the strongest connectivity, can be used to construct a projected Laplacian matrix $\mathcal{L}^{(k)}$, which serves as input for the $k$-means clustering step~\cite{ng2001spectral}. By performing $k$-means in a low-dimensional space, the algorithm ensures more efficient clustering of the data points. In the classical setting, computing all pairwise distances requires $\mathcal{O}(d N^2)$ time, while performing eigendecomposition on $\mathcal{L}$ scales as $\mathcal{O}(N^3)$, making the algorithm computationally prohibitive for large datasets.  

The quantum spectral clustering algorithm improves this by replacing the classical eigendecomposition with quantum linear-algebraic primitives (see Sec.~\ref{chapQML:subsec:qpca}). Under the assumption that Laplacian matrix can be efficiently stored in a QRAM structure, its eigenvalues and eigenvectors are efficiently obtained using QSVE or QSVT algorithms. Once the spectral embedding is obtained, clustering is performed on the resulting quantum states using the q-means algorithm, described earlier. Since both the spectral decomposition and the clustering subroutines benefit from asymptotic quantum speedups, the overall runtime is significantly reduced compared to the classical algorithm. Further theoretical details are available in Ref.~\cite{kerenidis2021quantum}.  

\end{mybox}

\subsection{Quantum reinforcement learning \label{chapQML:subsec:QRL} }
While supervised and unsupervised learning focus on identifying patterns, reinforcement learning (RL) is concerned with learning how to make a sequence of decisions that maximize an expected cumulative reward~\cite{sutton1998reinforcement}, and  is used in modern AI systems that rely on feedback-driven adaptation, including autonomous driving~\cite{kiran2021deep}, AlphaGo~\cite{silver2016mastering}, finance~\cite{hambly2023recent}, and drug discovery~\cite{popova2018deep}. 
Unlike the former paradigms, RL involves an interactive feedback loop between two main components: a decision-making agent and its task environment.  
At each step of interaction, the agent observes a state (or observation) $s \in \mathcal{S}$ from the environment and decides an action $a \in \mathcal{A}$ according to its current policy $\pi(a|s)$.  
Both the state and action spaces can be discrete or continuous, and the policy $\pi(a|s)$ governs the agent’s behavior, which can be either a deterministic mapping (or look-up table) or a stochastic probability distribution over actions.  

After taking an action $a_k$ at step $k$, the agent receives a scalar reward $r_k = R(s_k, a_k) \in \mathcal{R}$ determined by its action and transitions to a new state $s_{k+1}$ based on the environment dynamics.  
The state value function under policy $\pi$ is evaluated at each step as
\[
v_\pi(s) = \mathbb{E}_\pi\left[R_t \; \lvert \; s_t = s\right],~~~~ R_t = \sum_{k = 0}^\infty \gamma^k r_{t+k+1} \,,
\] 
where $R_t$ denotes the (expected) return in the future, and $\gamma \in [0, 1]$ a discount factor introduced to weight more the present reward and avoid divergence of the sum. Similarly, we can define Q-function (or state-action value function), which measures the expected return of taking action $a$ in state $s$ in the future for the given policy as
\[
Q_{\pi}(s, a) = \mathbb{E}_{\pi}[R_t  \; \lvert \; s_t = s, a_t = a]\;. 
\]
The goal of RL is to find the optimal policy $\pi^*$ that maximizes the expected return,
\[
\pi^* = \arg\max_\pi \mathbb{E}_\pi \!\left[\sum_{t=0}^\infty \gamma^t r_{t+1}\right] = \arg\max_a Q^*(s,a),
\]
where $Q^*(s,a) = \max_\pi Q_\pi(s,a)$ denotes the optimal Q-function for the highest achievable expected return.

RL algorithms are generally categorized into two main classes: value-based and policy-based methods.  Value-based methods, such as Q-learning~\cite{clifton2020q}, aim to learn the value function $Q(s,a; \boldsymbol{\theta})$ by iteratively updating its parameters $\boldsymbol{\theta}$ to approximate $Q^*(s,a)$, while policy-based methods, such as policy-gradient algorithms~\cite{sutton1999policy}, directly optimize a parameterized policy $\pi(a|s; \boldsymbol{\theta})$ using gradient-based approaches.  
A major challenge in RL arises from the combinatorial explosion of possible state–action pairs $(s,a)$.  
In real-world environments, the state and action spaces can be extremely large or continuous, making it infeasible to store or update $Q(s,a)$ for all combinations.  
To address this, deep neural networks are often used as function approximators to represent the policy or value functions, enabling generalization from previously explored states to unseen ones and scalability to high-dimensional tasks.  

Building on these classical foundations, quantum reinforcement learning (QRL) investigates how quantum computation can enhance learning efficiency, scalability, and representational power by capturing complex correlations between state and action spaces~\cite{meyer2022survey,lockwood2020reinforcement, chen2020variational, dunjko2015framework}.
A common QRL design uses PQCs to represent key components of the RL pipeline~\cite{skolik2023robustness}.  
In a value-based formulation, the Q-values can be represented as the expectation value of an observable $O_a$ corresponding to an action $a$, evaluated on quantum states generated by a unitary $U_{\boldsymbol{\theta}}(s)$ that encodes the current state~\cite{ lockwood2020reinforcement,chen2020variational, wu2025quantum, franz2023uncovering} as
\[
Q(s,a;\boldsymbol{\theta}) = \bra{0^{\otimes n}} U_{\boldsymbol{\theta}}^\dagger(s) O_a U_{\boldsymbol{\theta}}(s) \ket{0^{\otimes n}}.
\]
Here, the state $s$ may be encoded in the computational basis (for discrete state spaces) or through a continuous feature map, and the resulting expectation values correspond to Q-values for each action, enabling the agent to select actions with the highest expected rewards.  
In policy-based QRL, the policy $\pi(a|s; \boldsymbol{\theta})$ itself is implemented as a PQC that outputs a quantum state whose measurement statistics correspond to action probabilities~\cite{skolik2021quantum, sequeira2023policy, meyer2023quantum, jerbi2021parametrized}.  
Another approach leverages QBMs (see Sec.~\ref{chapQML:subsec:QGM}) as value-function approximators~\cite{crawford2016reinforcement, levit2017free, schenk2024hybrid}.  
In this setting, the Q-function can be, for instance, represented as an Ising-type Hamiltonian, where each state–action pair $(s,a)$ corresponds to an energy level used to approximate the Q-value.  
By tuning the system parameters, the Q-function is estimated as the expectation value of the Hamiltonian as $Q(s, a) = \langle H(s, a)\rangle$, typically evaluated via quantum annealing~\cite{ levit2017free,schenk2024hybrid}. As always, variational quantum RL techniques must be used with extreme care.

More generally, recent theoretical studies have established formal separations between classical and quantum RL, extending the quantum learnability results discussed in Sec.~\ref{chapQML:subsec:PAC} to interactive learning settings.  
For instance, under cryptographic hardness assumptions, there exist discrete environments (such as those defined by the discrete logarithm problem) where a quantum agent can achieve near-optimal policies that are provably intractable for classical learners within PAC framework~\cite{skolik2021quantum, jerbi2021parametrized}, leading to a separation between classical and quantum RL capabilities. 
Moreover, under the assumption of efficient quantum oracle access to the environment, QRL can achieve polynomial-to-exponential speedups in sample or query complexity compared to classical RL~\cite{dunjko2017advances, dunjko2016quantum}.  
In certain settings, quadratic improvements in learning time have also been demonstrated, reducing the required number of training iterations from $\mathcal{O}(T)$ in classical RL to $\mathcal{O}(\sqrt{T})$~\cite{wan2023quantum}, with experimental demonstrations of such speedups~\cite{saggio2021experimental}.  

Despite these promising results, most theoretical guarantees of quantum speedup in QRL rely on idealized assumptions, such as quantum-accessible environments~\cite{dunjko2017exponential}, where the agent can query the environment in superposition.  
While such assumptions enable rigorous complexity analysis, they certainly limit practical applicability. Identifying realistic QRL architectures that yield measurable quantum advantage remains an open and active area of research.

\section{The long reaching arm of quantum machine learning}

In this section we will briefly review some areas of QML that do not fit nicely in the previous sections, and which can roughly fall within the categories 2 and 3 of Fig.~\ref{chapQML:fig1}. Due to space constraints, we will purposely not cover important topics such as the use of ML for tasks such as enhancing quantum experiments~\cite{bukov2018reinforcement,niu2019universal,sivak2022model,madhusudhana2023optical,cimini2019calibration} (e.g., finding optimal shuttle routes for shuttling atoms in reconfigurable quantum computers~\cite{bluvstein2024logical}), decoding errors in fault-tolerant quantum devices~\cite{andreasson2019quantum,nautrup2019optimizing,bausch2024learning}, efficiently learning and classically representing states (via, e.g., neural quantum states~\cite{carleo2017solving,lange2024architectures} or tensor networks~\cite{shi2006classical,biamonte2017tensor,orus2014practical}) and their phase transitions~\cite{dong2019machine,rem2019identifying}, among others. Instead, we want to showcase two concrete examples where the tools that were originally developed for quantum learning have been used to: (i) solve tasks in quantum sensing, (ii) enhance how we solve quantum compilation, a learning task inherently classical. 

 In quantum sensing (quantum metrology), the task is to use a controllable probe quantum system to learn an unknown property of its environment. At its simplest level, the probe system is made to interact with the environment, which imprinted, e.g., some parameter of interest. A canonical example is magnetometry, where a system of qubits  accumulates a phase under the environment-induced Hamiltonian. That is, the system undergoes the transformation $\rho\rightarrow\rho(B)=e^{-i H(B) \tau}\rho e^{i H(B) \tau}$ for $H(B)=\sum_{j=1}^nBZ_j$, where $\tau$ is the  interrogation time, and $B$ the magnetic field of interest. Then, one seeks to estimate via measurements on the probe state the value of $B$~\cite{degen2017quantum,taylor2008high,rondin2014magnetometry}. Unlike the earlier sections, the relevant notion of advantage here is metrological rather than computational, which in a very real way eases the burden on the quantum protocol: classical simulability is a moot point, since a laptop simulation cannot replace an actual field measurement. What matters now are precision and resources, formalized by the (quantum) Cram\'er--Rao bound $\Delta^2 \hat{B}\,\ge\, [N F_B]^{-1}$, where $F_B$ is the (quantum) Fisher information per experiment, $N$ is the number of experimental repetitions and $\Delta^2\hat{B}$ the uncertainty in the estimated field $\hat{B}$~\cite{giovannetti2006quantum,pezze2018quantum,cerezo2021sub,sone2020generalized}.

Viewed through a sampling-complexity lens, the standard quantum limit corresponds to unentangled independent qubits, giving $F_B\in\mathcal{O}(1)$ per shot and $\Delta \hat{B}\in\mathcal{O}(N^{-1/2})$, hence to reach error $\varepsilon$ one needs $N\in\Theta(\varepsilon^{-2})$ samples. However, when the probe state is entanglement, one can obtain $F_B\in\mathcal{O}(\nu^2)$ in the ideal noiseless case (e.g., GHZ probes) (see~\cite{garcia2023effects, pezze2018quantum, huelga1997improvement,ijaz2024more} for an analysis of how noise affects the quantum Fisher information), achieving the Heisenberg limit  with $\Delta \hat{B}\in\mathcal{O}(N^{-1})$ and thus $N\in\Theta(\varepsilon^{-1})$ samples~\cite{giovannetti2006quantum,pezze2018quantum}. These ideas illustrate how a quantum sensing task can be readily framed as a learning problem. In particular, if we think about calibration as the ability to fine-tune the probe with evolutions $e^{-i H(B) \tau}$ for known $B$ and learning how the system's response (i.e., measurements on $\rho(B)$) change as a function of $B$, then the task can be naturally cast as a regression problem. This has lead to the use of moderm ML tools such as Bayesian regression~\cite{kaubruegger2021quantum,huerta2022inference},  variational techniques~\cite{cerezo2021sub,sone2020generalized,koczor2020variational,beckey2020variational,meyer2020variational} and Gaussian process regression~\cite{garcia2023deep,tsukamoto2022accurate}. More generally, beyond standard quantum sensors, it is expected that ML and QML techniques will be fundamental tools towards quantum perception~\cite{cerezo2022challenges}, facilitating the coherent capture, storage, and processing of sensor-generated quantum states in quantum memories and processors for posterior processing.

Next, let us turn our attention to quantum compiling. Quantum compiling broadly refers to mapping a high-level quantum algorithm to a device-native gate set that the hardware can implement. Typically one also seeks to minimize resource use, either to reduce the application of noisy gates in NISQ devices or to limit non-Clifford gates such as $T$-gates in stabilizer-code fault-tolerant architectures. While compiling can be performed coherently—often called quantum-assisted quantum compiling~\cite{khatri2019quantum,sharma2019noise}—we will focus on the common practice of compiling fully classically. Mathematically, compiling a unitary $U$ acting on a $d$-dimensional Hilbert space onto a trainable unitary $W_C$ can be cast as
\[
\min_{C\in\mathcal{V}^{\ast}} \; P\!\left(U, W_C\right)
\quad \text{s.t.} \quad \mathrm{cost}(C)\le \kappa,
\]
or, in penalized form, $
\min_{C\in\mathcal{V}^{\ast}} \; P\!\left(U, W_C\right) \;+\; \lambda\,\mathrm{cost}(C)$. Above, we have defined $\mathcal{V}=\{V_i\}_i$ is a fixed gate alphabet $\mathcal{V}^{\ast} \subseteq \mathcal{V}$ the gates one seeks to minimize, $W_C$ is the unitary implemented by circuit $C$ expressed as a products of gates in $\mathcal{V}$, and $P$ is a unitary/channel distance (e.g., $\|U-W_C\|$, or $\|\mathcal{U}-\mathcal{E}_C\|_{\diamond}$ in the noisy case with implemented channel $\mathcal{E}_C$). The term $\mathrm{cost}(C)$ penalizes resource usage—e.g., depth, two-qubit count, and the number of gates from the designated expensive subset $\mathcal{V}^{\ast} $—with $\lambda,\kappa$ encoding hardware constraints (connectivity, timing, error budgets). This viewpoint naturally recasts compiling as a learning problem, as one needs to train a circuits $C$ that minimize the above loss function.

Quantum compiling has been extensively tackled with tools from ML for adaptively discovering quantum circuits with variable architecture (e.g., via reinforcement learning)~\cite{tang2019qubit,grimsley2019adaptive,cincio2018learning,moro2021quantum,bilkis2021semi,du2020quantum,sim2021adaptive,zhang2021mutual,tkachenko2020correlation,claudino2020benchmarking,rattew2019domain,chivilikhin2020mog,zhang2020differentiable,wada2022sequential,li2024quarl,fosel2021quantum,nakaji2025quantum,rosenhahn2023monte}. However, more recently QML-specific tools have enhanced compilation--especially for time evolutions $U=e^{-iHt}$--by recasting the task as supervised learning from a dataset $\mathcal{S}=\{(\ket{\psi_i},\,U\ket{\psi_i})\}_{i=1}^N$. In practice, this requires that both the inputs $\{\ket{\psi_i}\}_i$ and the propagated states $\{U\ket{\psi_i}\}_i$ be efficiently representable on a classical computer, which quickly becomes intractable for generic choices due to the amount of entanglement accrued during the time evolution. A practical workaround is to compile a short-time propagator $U_{\delta t}=e^{-iH\delta t}$ with $\delta t=t/T$. For sufficiently small $\delta t$, $U_{\delta t}$ is close to the identity and does not generate large amounts of entanglement, so one can train a circuit $W_C$ for $U_{\delta t}$ and recover $U\approx (W_C)^T$. The key difficulty is entanglement growth across slices: while the first application of $U_{\delta t}$ can be trained on product or low-entangled inputs that admit tensor-network representations, the intermediate states $U_{\delta t}^k\ket{\psi_i}$ become progressively more entangled, making it harder to generate training pairs and to evaluate losses classically. This issue forces a careful choice of the training ensemble and schedule. Recently, it was shown that one can actually selects low-entanglement (often even tensor product) inputs and short times so that both $\ket{\psi_i}$ and $U_{\delta t}\ket{\psi_i}$ remain tractable via tensor networks, and then leverages out-of-distribution generalization guarantees that were derived within the context of QML~\cite{caro2021generalization,caro2022outofdistribution}. This notable result allows one to effectively learn dynamics and  extend performance beyond the training distribution: train on simple tensor product states, generalize to arbitrary ones~\cite{kukliansky2024leveraging,zhang2024scalable}.

\section{Conclusion}

QML finds itself at an exciting yet knife’s-edge moment, with its ultimate capabilities for practical tasks still uncertain and actively debated. On the one hand, formal separations between quantum and classical learners do exist, although they often arise in oracle or cryptographic regimes and are not yet pathways to broadly useful algorithms. On the other hand, approaches for bridging the gap between these separations and day-to-day problems have hit serious roadblocks. While there is likely no single cause for such impasses, this review’s organizing view highlights that most quantum learning proposals for analyzing classical data fall into two broad strands.

The first builds learning models directly from flagship quantum algorithms. This “QML-from-the-best-quantum” approach drove the early-2010s push to leverage primitives such as QRAM and the HHL algorithm to obtain provable speedups on tasks like qPCA or recommendation systems. Much of this promise was later dequantized. The second strand, “QML-from-the-best-classical,” led to variational QML, which takes its cue from modern classical ML and seeks to implement deep-learning methodology at the quantum level. Despite encouraging initial heuristics, variational QML faces well-documented issues of trainability and scalability, and recent work indicates that critical aspects of its information-processing power admit classical surrogates and effective dequantizations. Together, these categories hint at a useful lesson: Rather than borrowing from the “best” of quantum and classical, we may need simpler, more interpretable quantum pipelines whose assumptions and data-access models are explicit and auditable, and whose performance can be stress-tested against strong classical baselines.

By contrast, for quantum data the picture is already brighter. QML has sharpened how we learn properties of states and channels, design measurements, compress information via randomized protocols, and close the loop between experiment and inference. Here the data are intrinsically quantum, access models are physically grounded, and quantum methods have delivered provable and practical improvements over classical post-processing that are widely used across laboratories in academia and industry. Whatever happens for classical tasks, these quantum-for-quantum successes will remain.

Indeed, the success of QML for quantum data--which intrinsically follows the principles of quantum mechanics--suggests a natural role for quantum-native generative modeling based on the pipeline of state preparation, sampling, and learning via measurements. Of course, progress here still relies on not overselling near term benefits and on being precise about where sampling structure or physical priors truly help. These lines favor modest, well-specified quantum advantages over sweeping claims.

Overall, QML is going through a lively and fast-moving area. Progress will come from studying the interplay among data models, encodings, expressivity, landscapes, generalization, noise, resource accounting as well from finding new areas where the theoretical principles of QML can be applied. For the readers coming into the field, we can only but advocate careful benchmarking, comparisons against classical surrogates, and reporting negative results alongside positive ones. Enthusiasm is warranted, but so is discipline: separate signal from hype, prefer clear assumptions to sweeping narratives, and follow the evidence wherever it leads.

\section{Acknowledgments}
We thank I-Chi Chen for useful comments and feedback regarding our manuscript. SYC was supported by the U.S. Department of Energy (DOE), Office of Science, Office of Advanced Scientific Computing Research under Contract No. DE-AC05-
00OR22725 through the Accelerated Research in Quatum Computing Program MACH-Q project. MC acknowledges support by the Laboratory Directed Research and Development (LDRD) program of Los Alamos National Laboratory (LANL) under project number 20260043DR and by LANL ASC Beyond Moore’s Law project. This material is based upon work supported by the U.S. Department of Energy, Office of Science, National Quantum Information Science Research Centers, Quantum Science Center (LC).

\bibliography{quantum}

\end{document}